\pdfoutput=1
\documentclass[a4paper,11pt]{article}
\usepackage[margin=2.5cm]{geometry}
\usepackage[utf8]{inputenc}
\usepackage[T1]{fontenc}
\usepackage{hyperref}
\usepackage{amsmath,amsfonts,bbm,nicefrac}
\usepackage{tikz}
\usepackage{caption}

\graphicspath{{figpdf/}}

\makeatletter

\@addtoreset{equation}{section}
\makeatother

\newcommand{\secref}[1]{Sec.~\ref{sec:#1}}
\newcommand{\figref}[1]{Fig.~\ref{fig:#1}}

\newcommand{\normal}[1]{\ :\! #1 \!\!:}
\newcommand{\aver}[1]{\left\langle {#1} \right\rangle}

\newcommand{\ket}[1]{| {#1} \rangle}
\newcommand{\vect}[1]{\boldsymbol{#1}}
\newcommand{\W}[5]{W \left( \!\!\left.\begin{array}{cc} #1&#2 \\ #4&#3 \end{array}\right|#5 \right)}

\newcommand{\id}{\mathbbm{1}}

\newcommand{\nt}{\widetilde{n}}
\newcommand{\nn}{\nonumber}
\newcommand{\wh}{\widehat}
\newcommand{\wt}{\widetilde}
\newcommand{\compi}{\mathrm{i}}

\newcommand{\Tr}{\mathop{\mathrm{Tr}}}

\newcommand{\looppict}[1]{\scalebox{0.2}{
    \begin{tikzpicture}[line cap = round, line join = round,ultra thick, teal]
      \node (1) at (0,0){};
      \node (2) at (1,0){};
      \node (3) at (2,0){};
      \node (4) at (3,0){};
      \node (5) at (4,0){};
      \node (6) at (5,0){};
      \node (extl1) at (-0.5,-0.5) {};
      \node (extl2) at (-0.5,-1) {};
      \node (extr1) at (5.5,-0.5) {};
      \node (extr2) at (5.5,-1) {};
      \fill (1) circle [radius=4pt];
      \fill (2) circle [radius=4pt];
      \fill (3) circle [radius=4pt];
      \fill (4) circle [radius=4pt];
      \fill (5) circle [radius=4pt];
      \fill (6) circle [radius=4pt];
      #1
    \end{tikzpicture}}
}

\title{The fully packed loop model as a non-rational $W_3$ conformal field theory}
\author{T. Dupic, B. Estienne, Y. Ikhlef}

\begin{document}

\maketitle

\begin{abstract}
The fully packed loop (FPL) model is a statistical model related to the integrable $U_q(\wh{\mathfrak{sl}}_3)$ vertex model. In this paper we study the continuum limit of the FPL. With the appropriate weight of non-contractible loops, we give evidence of an extended $W_3$ symmetry in the continuum. The partition function on the torus is calculated exactly, yielding new modular invariants of $W_3$ characters. The full CFT spectrum is obtained, and is found to be in excellent agreement with exact diagonalisation.
\end{abstract}

\section{Introduction}
\label{sec:intro}

Loop models are lattice statistical models with non-local Boltzmann weights, which generally describe extended geometrical objects such as spin interfaces in the Ising model, or percolation clusters. Since the introduction of the Coulomb Gas approach \cite{Nienhuis_82,DotsenkoFateev84}, they have been identified as exactly solvable lattice realizations of non-rational Conformal Field Theories (CFTs) with a generic central charge. The most studied example is the critical O($n$) loop model, related to the six-vertex model and the Temperley-Lieb algebra on the lattice, and to the Virasoro algebra in the continuum. The critical regime $-2<n \leq 2$ corresponds to a central charge $-\infty<c \leq 1$. The bulk excitation spectrum of the O($n$) loop model can be described in terms of an infinitely extended Kac table, which has led to many physical results in the early stages of CFT (see \cite{Nienhuis84}): magnetic exponents and fractal dimension of interfaces in spin models, connectivity exponents in percolation, etc. More recently, some remarkable progress was made in the study of the operator algebra of the O($n$) model \cite{DelfinoViti11,Picco-etal13}, in particular the computation of various classes of coefficients of the Operator Product Expansion (OPE) of primary operators \cite{EstienneIkhlef15}. In this context, interesting connections have been found between the O($n$) model and the time-like Liouville theory  \cite{DelfinoViti11,Picco-etal13,EstienneIkhlef15,Ikh3points16}.

In this paper, we consider a variant of the O($n$) loop model, namely the Fully Packed Loop (FPL) model on the hexagonal lattice. This model was introduced by Reshetikhin \cite{reshetikhin91}, and is related to the fifteen-vertex model with $U_q(\wh{\mathfrak{sl}}_3)$ symmetry. The spectrum of the corresponding quantum spin chain has been studied numerically \cite{AlcarazMartins90} and analytically \cite{PZJ98}. On one hand, when appropriate twisted periodic boundary conditions (PBC) are applied, it was shown that the $U_q(\wh{\mathfrak{sl}}_3)$ spin chain scales to the $W_3$ algebra \cite{Fateev87}, an extension of the Virasoro algebra including a spin-3 conserved current $W(z)$, additionally to the stress-energy tensor $T(z)$. On the other hand, the Coulomb Gas formalism for the FPL model was developed in \cite{kondev1996operator}. In both approaches, the scaling theory was found to be a CFT with central charge in the range $-\infty<c \leq 2$. The aim of the present work is to make a connection between the Coulomb Gas construction of \cite{kondev1996operator} and the free-field realization of the $W_3$ algebra \cite{Fateev87}. This opens the way to a complete description of the excitation spectrum of the FPL model in terms of an extended $W_3$ Kac table, and allows us to exhibit an interesting class of modular invariant partition function, possibly related but different from the classified $W_3$ modular invariants \cite{Gannon94}.

The structure of the paper is the following. In \secref{models}, we recall the definition of the FPL model, and its relation with the integrable $U_q(\wh{\mathfrak{sl}}_3)$ vertex and face models. In \secref{W3}, we study the continuum limit of the FPL model when the loop fugacities for contractible and non-contractible (i.e. topologically non-trivial) loops are respectively of the form $n=2\cos\lambda$ and $\nt=2\cos2\lambda$, with $0\leq\lambda<\pi$. We can then classify the spectrum of the loop model in terms of primary operators from an extended $W_3$ Kac table, and their descendants under the $W_3$ algebra. These results are verified numerically by an exact diagonalisation procedure, explained in \secref{numerics}. In \secref{torus}, we turn to the study of toroidal partition functions: by applying the steps of \cite{difrancesco_saleur_zuber_coulomb}, we derive the full spectrum of conformal dimensions of the FPL model, and express the FPL model partition function on the torus in terms of Coulombic partition functions. Particular values of the loop fugacity $n=\sqrt{2}$ and $n=1$ are examined in more detail. The Appendix contains some technical calculations needed in~\secref{torus}.

\section{The Fully-Packed Loop model and related lattice models}
\label{sec:models}

\subsection{The  loop model}
\label{sec:loop_model}

\begin{figure}
  \centering
  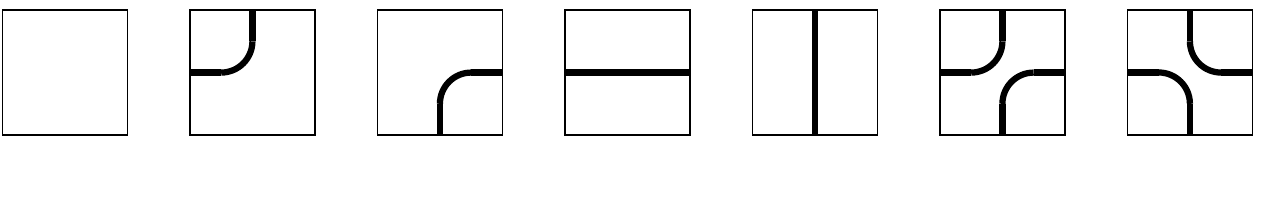 \\
  \smallskip (a) \\
  \vspace{1cm}
  \includegraphics[scale=0.5]{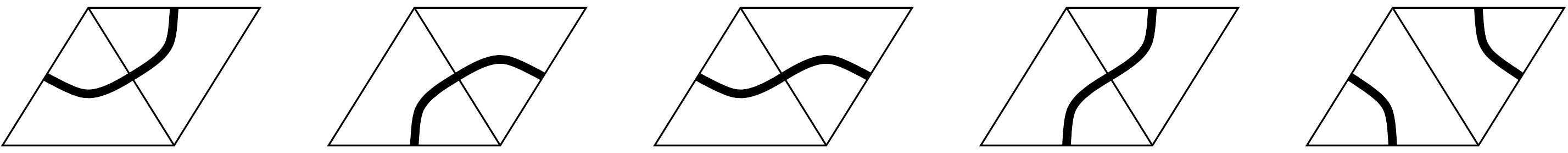}  \\
  \smallskip (b)
  \caption{(a) The plaquette configurations of the $A_2^{(1)}$ loop model on the square lattice. (b) The double-plaquette configurations of the Fully Packed Loop model on the hexagonal lattice, obtained by setting $a_1=a_6=0$ and $a_2=a_3=a_4=a_5=a_7$.}
  \label{fig:loop}
\end{figure}

The $A_2^{(1)}$ loop model on the square lattice is defined in \cite{warnaar93} as follows. The allowed configurations are those represented in \figref{loop} a, their local weights are given in terms of the crossing parameter $\lambda$ and the spectral parameter $u$ :
\begin{equation}
  \label{eq:weight_a21_loop}
  a_1, \dots a_7 = \sin(\lambda - u) , e^{+iu}\sin\lambda , e^{-iu}\sin\lambda , \sin u , \sin u , \sin(\lambda - u) , \sin u \,,  
\end{equation}
and each closed loop gets a weight $n=2\cos\lambda$. On a surface with cycles (cylinder, torus \dots), the non-trivial loops (those which wind around the cycles of the surface) get a different weight $\wt n$.
Hence, the partition function reads:
\begin{equation}
  \label{eq:partition_function_a21_loop}
  Z_{\text{loop}}(n,\wt n,u) = \sum_{\text{loop config.}\ C} a_i^{N_i(C)}
  n^{L(C)} \ \wt{n}^{\wt{L}(C)} \,,
\end{equation}
where the sum is over every loop configuration $C$ on the square lattice obtained by combining the plaquettes $\{a_1, \dots a_7\}$, $N_i(C)$ is the number of plaquettes of type $a_i$ appearing in $C$, and $L(C)$ [resp. $\wt L(C)$] is the number of trivial (resp. non-trivial) closed loops in $C$. Also, note that $Z_{\text{FPL}}(n=\wt n=2)$ is equal to the number of three-colourings of the hexagonal lattice \cite{kondev1996operator}: if the colours are called $(A,B,C)$, each empty edge is labelled $A$, and the edges along a closed loop are labelled $BC\!BC\!BC\dots$ or $C\!BC\!BC\!B\dots$, resulting in loop fugacities $n=\wt n=2$.

For a generic value of $u$, the correct embedding of the square lattice is given by rhombi of the form
\begin{center}
  \begin{picture}(0,0)%
\includegraphics{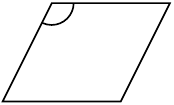}%
\end{picture}%
\setlength{\unitlength}{4144sp}%
\begingroup\makeatletter\ifx\SetFigFont\undefined%
\gdef\SetFigFont#1#2#3#4#5{%
  \reset@font\fontsize{#1}{#2pt}%
  \fontfamily{#3}\fontseries{#4}\fontshape{#5}%
  \selectfont}%
\fi\endgroup%
\begin{picture}(789,474)(574,-1693)
\put(856,-1501){\makebox(0,0)[lb]{\smash{{\SetFigFont{12}{14.4}{\familydefault}{\mddefault}{\updefault}{\color[rgb]{0,0,0}$\theta$}%
}}}}
\end{picture}%

\end{center}
with opening angle $\theta = \frac{2}{3} \pi u/\lambda$.

The ``isotropic'' point sits at the value $u = \lambda$ : for this value, one has, after dropping the irrelevant factors $e^{\pm i u}$ :
\begin{equation}
  a_1 = a_6 = 0 \,,
  \qquad
  a_2=a_3=a_4=a_5=a_7=\sin\lambda \,.
\end{equation}
At this value $u = \lambda$, the $A_2^{(1)}$ loop model reduces \cite{reshetikhin91} to the fully-packed loop  (FPL) model on the hexagonal lattice (see \figref{loop} b). The partition function becomes
\begin{align}
  Z_{\text{loop}}(n,\wt n,u=\lambda) &= (\sin\lambda)^{\mathcal{N}/2} \sum_{\text{FPL loop config.}\ C}
  n^{L(C)} \ \wt{n}^{\wt{L}(C)} \nn \\
  &\equiv (\sin\lambda)^{\mathcal{N}/2} \times Z_{\text{FPL}}(n,\wt n) \,,
  \label{eq:partition_function_loop}
\end{align}
where the sum is over every loop configuration $C$ on the hexagonal lattice visiting each vertex exactly once, and $\cal N$ is the total number of vertices.

\subsection{The fifteen-vertex model}
\label{sec:15V}

\begin{figure}
  \centering
  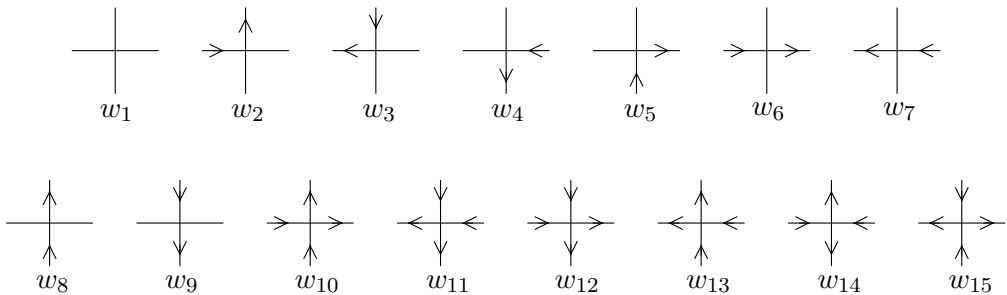
  \caption{Configurations of the $15$-vertex model}
  \label{fig:conf_15_vertex}
\end{figure}

The 15-vertex model is defined by the $R$-matrix acting on two fundamental representations of the $U_q(\wh{\mathfrak{sl}}_3)$ quantum affine algebra, whose basis is taken as $( \ket{\uparrow}, \ket{\cdot}, \ket{\downarrow})$. The corresponding vertex configurations are shown in \figref{conf_15_vertex}, and their weights are given by \cite{jimbo1986}:
\begin{equation}
  \label{eq:weights_A21_vertex}
  \begin{split}
    & w_1 = w_{10} = w_{11} = \sin(\lambda - u) \,, \\
    & w_2 = w_4 = w_{14} = e^{+ \compi u } \sin \lambda \,, \\ 
    & w_3 = w_5 = w_{15} = e^{- \compi u } \sin \lambda \,, \\
    & w_6 = w_7 = w_8 = w_9 = w_{12} = w_{13} = \sin u \,. \\
  \end{split}
\end{equation}

When spectral parameters are attached to the lines of the vertex model, we use the following graphical conventions:
\begin{equation*}
  R(u_1-u_2) = \qquad \raisebox{-1.2cm}{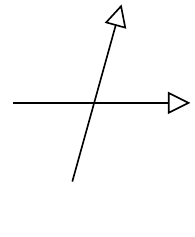} \quad.
\end{equation*}

The 15-vertex model can be obtained in a direct way from the loop model, by using the trick of \cite{Baxter-book}. Let us describe this equivalence on the plane first, where no non-trivial loop can occur. One gives an orientation to each loop (independently of the others), and assigns a weight $e^{i\lambda}$ (resp. $e^{-i\lambda}$) to the anti-clockwise (resp. clockwise) oriented loops, so that the total loop weight is indeed $n=2\cos\lambda$. Then, the phase factor for each loop is distributed locally, by attributing a factor $\exp [i\alpha \lambda/(2\pi)]$ to each oriented loop segment in \figref{loop} which turns by an angle $\alpha$ (we take the convention that the loop segments cross the edges of the rhombic plaquettes orthogonally). Then, on each edge carrying the spectral parameter $u$, one inserts the operator $\id = e^{+iu \eta} \times e^{-iu \eta}$, where $\eta = \mathrm{diag}(\frac{1}{3},1,-\frac{1}{3})$, so that the factor $e^{+iu \eta}$ (resp. $e^{-iu \eta}$) acts on the incoming (resp. outgoing) vector space of the adjacent R-matrix. This last step may be viewed as a change of gauge. Through the whole procedure, the loop model with weights \eqref{eq:weight_a21_loop} maps to the 15-vertex model with weights \eqref{eq:weights_A21_vertex}.

On surfaces with cycles, the non-trivial loops have a total winding equal to zero, and must be treated separately. On a cylinder of width $N$ sites and circumference $M$ sites, the partition function of the loop model is obtained by introducing twisted boundary conditions:
\begin{equation}
  Z_{\mathrm{cyl}}(n, \nt) = \mathrm{tr} \left[ (K_\mu \otimes \dots \otimes K_\mu) (t_N)^M \right] \,,
  \qquad \nt=2\cos\mu \,,
\end{equation}
where $t_N$ is the row-to-row transfer matrix, $\mathrm{tr}$ stands for the conventional trace, and $K_\mu = \mathrm{diag}(e^{i\mu},1,e^{-i\mu})$.

On the torus, the weight of non-trivial loops cannot be distributed locally into the vertex model: rather, to give non-trivial loops a weight $\nt=2\cos\mu$, each vertex configuration with arrow fluxes $m$ and $m'$ through the two cycles of the torus must be given the non-local phase factor $\exp [i\mu (m \wedge m')]$, where $m \wedge m'$ denotes the greatest common divisor (gcd) of $m$ and $m'$.


\subsection{The RSOS model}
\label{sec:RSOS}

\begin{figure}
  \centering
  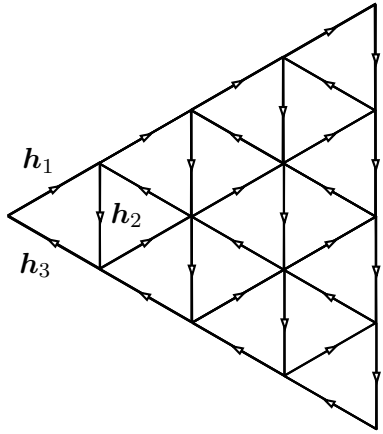
  \caption{The height graph $A_{k=4}$.}
  \label{fig:graph_Ak}
\end{figure}

The third model of interest is the Restricted Solid-On-Solid (RSOS) model based on the Weyl alcove $A_k$ of $\mathfrak{sl}_3$ \cite{Jimbo1987}. Let us first set up our conventions for $\mathfrak{sl}_3$, which shall be used throughout the paper. The simple roots $(\vect e_1, \vect e_2 )$ are two-dimensional vectors with scalar products:
\begin{equation}
  \vect e_1 \cdot \vect e_1 = \vect e_2 \cdot \vect e_2 = 2 \,,
  \qquad
  \vect e_1 \cdot \vect e_2 = -1 \,.
\end{equation}
The fundamental weights $(\vect\omega_1,\vect\omega_2)$ are given by:
\begin{equation}
  \vect\omega_1 = \frac{1}{3}(2\vect e_1+\vect e_2) \,,
  \qquad \vect\omega_2 = \frac{1}{3}(2\vect e_2+\vect e_1)
  \qquad \Leftrightarrow
  \qquad\vect\omega_i \cdot \vect e_j = \delta_{ij} \,.
\end{equation}
This gives $\vect\omega_1 \cdot \vect\omega_1= \vect\omega_2 \cdot \vect\omega_2=2/3$ and $\vect\omega_1\cdot \vect\omega_2=1/3$. The root and weight lattices are respectively:
\begin{equation} \label{eq:R}
  \mathcal{R} = \mathbb{Z} \vect e_1 + \mathbb{Z} \vect e_2 \,,
  \qquad
  \mathcal{R}^* = \mathbb{Z} \vect\omega_1 + \mathbb{Z}\vect\omega_2 \,.
\end{equation}
The weights of the fundamental representation are
\begin{equation}
  \vect h_1 =\vect\omega_1 \,,
  \qquad \vect h_2 = \vect\omega_2-\vect\omega_1 \,,
  \qquad \vect h_3 = -\vect\omega_2 \,.
\end{equation}
The Weyl vector is $\vect\rho=\vect e_1 + \vect e_2 = \vect\omega_1+\vect\omega_2$.
The oriented graph $A_k$ is defined as follows (see also \figref{graph_Ak}). The set of vertices of $A_k$ is given by  the dominant integral weights of  $\wh{\mathfrak{su}}(3)_k$, namely 
\begin{align}
  A_k = \{ \lambda_1 \vect\omega_1 + \lambda_2 \vect\omega_2 , \, \lambda_i \in \mathbb{N} , \, \lambda_1+\lambda_2  \leq  k \} \,,
\end{align}
and the edges of $A_k$ are oriented along the three vectors $(\vect h_1,\vect h_2,\vect h_3)$. In the following, we will refer to $A_k$ as {\it the height graph}.

In the RSOS model, each vertex of the square lattice carries a height variable which is a vertex of $A_k$, and the Boltzmann weight of a face is denoted by:
\begin{equation}
  \W {\vect a}{\vect b}{\vect c}{\vect d}u
  = \quad \raisebox{-1cm}{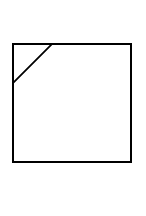} 
  = \qquad \raisebox{-1cm}{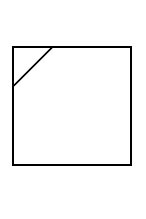} 
  \quad= \W{\vect a}{\vect a+\vect h_\kappa}{\vect a+\vect h_\mu+\vect h_\nu}{\vect a+\vect h_\mu}{u} \,,
\end{equation}
where the labels $1 \leq \kappa,\sigma,\nu,\mu \leq 3$, and must satisfy $\vect h_\mu+\vect h_\nu=\vect h_\kappa+\vect h_\sigma$. Setting $\mu \neq \nu$, the face weights are given by:
\begin{equation}
  \begin{aligned}
    \raisebox{-1cm}{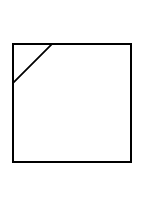} \quad &= \sin(\lambda-u) \,, \\
    \raisebox{-1cm}{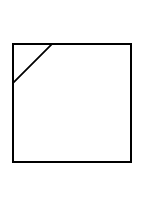} \quad &= \frac{\sin\lambda \ \sin(\lambda a_{\mu\nu}+u)}{\sin\lambda\, a_{\mu\nu}} \,, \\
    \raisebox{-1cm}{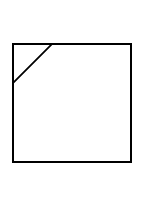} \quad &= \frac{\sin u \sin \lambda(a_{\mu\nu}+1)}{\sin \lambda\, a_{\mu\nu}} \,,
  \end{aligned}
\end{equation}
where $\lambda \equiv \pi/(k+3)$ and $a_{\mu\nu} \equiv (\vect a+\vect\rho) \cdot (\vect h_\mu-\vect h_\nu)$.
On simply connected domains, the RSOS model is related to the 15-vertex model through the vertex-face correspondence~\cite{JMO88}. On the cylinder, the relation between these two models (along with the FPL model) uses the Hecke algebra symmetry, combined with the Markov trace: see next section.

\subsection{Hecke algebra and Markov trace}
\label{sec:Hecke}

The FPL, 15-vertex and RSOS models are three realizations of the $\mathfrak{su}(3)$ Hecke algebra. In all three cases, the Boltzmann weight of a square plaquette at position $j$ on the lattice can be written as
\begin{equation}
  \sin(\lambda - u)\, \id + \sin u \, U_j \,,
\end{equation}
where $U_1, U_2 \dots U_{N-1}$ are the Hecke generators, obeying the algebraic relations:
\begin{align}
  &U_j^2 = 2\cos\lambda \times U_j \,, \label{eq:Hecke1} \\
  &U_j U_{j+1}U_j - U_j = U_{j+1}U_jU_{j+1} - U_{j+1} \,, \label{eq:Hecke2}  \\ 
  &U_{j'}U_j = U_jU_{j'} \qquad\qquad \text{if } |j-j'| > 1 \,, \label{eq:Hecke3}  \\
  &(U_{j-1} - U_{j+1} U_j U_{j-1} + U_j)(U_j U_{j+1} U_j - U_j) = 0 \,.
  \label{eq:vanishing_symmetryzer}
\end{align}
Equations (\ref{eq:Hecke1}--\ref{eq:Hecke3}) are the defining relations of the Hecke algebra, and the additional relation~\eqref{eq:vanishing_symmetryzer} defines the $\mathfrak{su}(3)$ quotient of the Hecke algebra, which we denote as $\mathcal{H}_N$ in the following.

When the model is defined on a cylinder of $N \times M$ sites (where $M$ is the circumference), the partition function is defined as
\begin{equation}
  Z_{\mathrm{cyl}} = \Tr \left[(t_N)^M \right] \,,
\end{equation}
where $t_N$ is the row-to-row transfer matrix with open boundary conditions, and $\Tr$ is a linear form on $\mathcal{H}_N$, which obeys the {\it Markov property} :
\begin{equation} \label{eq:Markov}
  \forall x \in \mathcal{H}_j \,, \qquad
  \Tr(U_j\, x) = \frac{\sin3\lambda}{\sin2\lambda} \times \Tr x \,,
\end{equation}
where $\mathcal{H}_j$ is the sub-algebra generated by $\{ U_1, \dots U_{j-1} \}$.
A linear form obeying this property is called a Markov trace. Let us give the explicit form of the generators $U_j$ and the Markov trace $\Tr$ for the three lattice models of interest.

\begin{itemize}

\item In the FPL model, the Hecke generator takes the form
  \begin{equation} \label{eq:Uj-loop}
    U = \qquad \raisebox{-0.55cm}{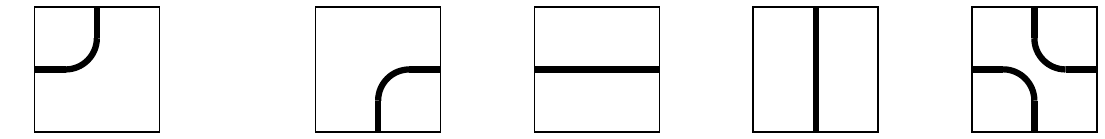} \quad.
  \end{equation}
  The Markov trace is $\Tr x = (2\cos2\lambda)^{\wt{L}(x)}$, where $\wt{L}(x)$ is the number of closed loops appearing when the top and bottom part of the diagram $x$ are identified. In other words, the Markov property is obeyed when $\nt=2\cos 2\lambda$.

\item In the 15V model, the matrix $\check R(u) \equiv PR(u)$, where $P$ is the permutation operator, has the form $\check R_{j,j+1}(u)=\sin(\lambda - u)\, \id + \sin u \, U_j$, with
  \begin{equation}
    U = \left(\begin{array}{ccccccccc}
        0 & 0 & 0 & 0 & 0 & 0 & 0 & 0 & 0 \\
        0 & e^{+i\lambda} & 0 & 1 & 0 & 0 & 0 & 0 & 0 \\
        0 & 0 & e^{+i\lambda} & 0 & 0 & 0 & 1 & 0 & 0 \\
        0 & 1 & 0 & e^{-i\lambda} & 0 & 0 & 0 & 0 & 0 \\
        0 & 0 & 0 & 0 & 0 & 0 & 0 & 0 & 0 \\
        0 & 0 & 0 & 0 & 0 & e^{+i\lambda} & 0 & 1 & 0 \\
        0 & 0 & 1 & 0 & 0 & 0 & e^{-i\lambda} & 0 & 0 \\
        0 & 0 & 0 & 0 & 0 & 1 & 0 & e^{-i\lambda} & 0 \\
        0 & 0 & 0 & 0 & 0 & 0 & 0 & 0 & 0
      \end{array}\right) \,,
  \end{equation}
  expressed in the basis $( \ket{\uparrow}, \ket{\cdot}, \ket{\downarrow}) \otimes ( \ket{\uparrow}, \ket{\cdot}, \ket{\downarrow})$. The Markov trace is given by
  \begin{equation}
    \Tr x = \mathrm{tr}[K^{\otimes N} x] \,,
  \end{equation}
  where $\mathrm{tr}$ denotes the conventional trace, and $K=\mathrm{diag}(e^{2i\lambda},1,e^{-2i\lambda})$.

\item In the RSOS model, the face weights can be written
  \begin{equation}
    \W {\vect a}{\vect b}{\vect c}{\vect d}u = \sin(\lambda-u) \  \delta_{\vect b,\vect d} +
    \sin u \  U \left(\begin{array}{cc} \vect a & \vect b \\ \vect d & \vect c \end{array}\right) \,,
  \end{equation}
  where the Hecke generators read:
  \begin{equation}
    U \left(\begin{array}{cc} \vect a & \vect a+\vect h_\kappa \\ \vect a+\vect h_\mu & \vect a+\vect h_\mu+\vect h_\nu \end{array}\right)
    =  (1-\delta_{\mu\nu}) \frac{\sin\lambda(a_{\mu\nu}+1)}{\sin\lambda a_{\mu\nu}} \,.
  \end{equation}
  The Markov trace in this model is given in the Appendix of~\cite{pasquier_sun_rsos}.
\end{itemize}

As it is argued in~\cite{pasquier_sun_rsos}, the algebraic relations (\ref{eq:Hecke1}--\ref{eq:Hecke3}) and the Markov property \eqref{eq:Markov} determine completely the value of $\Tr x$ for any $x \in \mathcal{H}_N$. Hence, the FPL, 15-vertex and RSOS models have the same partition function on the cylinder.

\section{Continuum limit and $W_3$ algebra}
\label{sec:W3}

The lattice models from the previous section are all critical, and are described in the continuum by a conformal field theory with an extended $W_3$ symmetry. This $W_3$ algebra is a chiral algebra generated by two fields \cite{Zamolodchikov85,Fateev87}: the spin $2$ stress energy tensor, and an additional spin $3$ field $W(z)$. The modes of these two chiral fields enjoy the following commutation relations
\begin{equation}
  \label{eq:W3_relations}
  \begin{split}
    \left[L_n, L_m \right] &= (n-m) L_{n+m} + \frac{c}{12}(n^3 - n) \delta_{n+m,0}\\
    \left[L_n, W_m \right] &= (2n - m) W_{n+m}\\
    \left[W_n, W_m \right] &= \frac{c}{360} (n^2 - 4) (n^2 - 1) n \delta_{n+m,0} + \frac{16}{22+ 5 c} (n-m) \Lambda_{n+m} \\  
    & \quad +  \frac{1}{30}(n-m) \left( 2m^2 + 2 n^2 - mn - 8 \right) L_{n+m} \,,
  \end{split}
\end{equation}
where $\Lambda_n$ are the modes of the (quasi-primary) composite field
$\Lambda(z) = \normal{T^2(z)} - \frac{3}{10} \partial^2 T(z)$, namely 
\begin{align}
  \Lambda_n = \sum_{k= -\infty}^{+\infty} \normal{L_{k} L_{n-k}} + \frac{\left(1 + \left\lceil \frac{n}{2} \right \rceil \right)\left(1 + \left\lfloor \frac{n}{2} \right \rfloor \right)}{5} L_n
\end{align}
In this section we describe several CFTs with $W_3$ symmetry. We start with the standard and simplest case, namely the $W_3$ minimal models, which describe the continuum limit of the RSOS model described in the previous section \cite{pasquier_sun_rsos}. We then move on to the more subtle case of the loop model, by means of a Coulomb gas argument. Finally we mention the (time-like) Toda field theory, which we conjecture to describe the purely electric sector of the loop model.

\subsection{Minimal models $W_3(p,q)$}
\label{sec:minimal}

Throughout this section, we use the conventions of \secref{RSOS} for $\mathfrak{su}(3)$ roots and weights.  The minimal models $W_3(p,q)$ -- with $p$ and $q$ coprime integers, $p,q \geq 3$  -- are rational conformal field theories with $W_3$ symmetry and central charge
\begin{align}
  c(p,q) = 2 \left(1 - 12 \frac{(q-p)^2}{p q} \right)
\end{align} 
They have finitely many scalar\footnotemark \footnotetext{We are concerned with the simplest $W_3$ minimal models, \emph{i.e.} those with diagonal modular invariant \cite{Beltaos2012}.} primary fields $\Phi_{\vect{n},\vect{m}}$ , with $\vect{n} \in \vect{\rho}  + A_{p-3}$ and $\vect{m}  \in \vect{\rho}  + A_{q-3}$, \emph{i.e.}
\begin{equation}
  \begin{aligned}
    \vect{n} &\in \{ n_1 \vect\omega_1 + n_2 \vect\omega_2, \, n_i \geq 1, \, n_1 + n_2 \leq p-1 \},\\ 
    \vect{m} &\in \{ m_1 \vect\omega_1 + m_2 \vect\omega_2, \, m_i \geq 1, \, m_1 + m_2 \leq q-1 \} \,.
  \end{aligned}
\end{equation}
The conformal dimension of $\Phi_{\vect{n},\vect{m}}$ is 
\begin{align}
  \Delta_{\vect{n},\vect{m}} = \bar{\Delta}_{\vect{n},\vect{m}} = \frac{\left(q \vect{n} -p \vect{m} \right)^2 -  2(q-p)^2}{2pq} \,.
\end{align}
The minimal models $W_3(p,q)$ can be mapped to a two component Coulomb gas, \emph{i.e.} a bosonic field  $\vect{\varphi}(z) = (\varphi_1(z),\varphi_2(z))$  in the presence of a background charge $ \vect\alpha_0$ with
\begin{equation}
  \vect\alpha_0 = (\alpha_+ + \alpha_-) \vect{\rho} \,,
  \qquad \alpha_+ = \sqrt{q/p} \,,
  \qquad \alpha_- = - \sqrt{p/q} \,.
\end{equation}
Note that the vector $\vect\rho=\vect\omega_1+\vect\omega_2$ has square norm $\vect\rho^2=2$.
The stress-energy tensor and central charge are 
\begin{align}
  T(z) = - \frac{1}{2} :  ( \partial \vect\varphi )^2 : + i \vect\alpha_0 \cdot \partial^2 \vect\varphi \,,
  \qquad c = 2 - 12 \, \vect\alpha_0^2 \,,
\end{align}
and the bosonic field is normalized as  $\aver{\varphi_i(z) \varphi_j(0)} = - \delta_{ij} \ln z$. The bosonic expression of the spin 3 field $W$ is  a bit more involved, and can be found in  \cite{Fateev87}. Vertex operators
$V_{\vect{\alpha}}(z) = \normal{e^{\compi \vect\alpha \cdot \vect\varphi(z)}}$
are primary fields w.r.t. the extended $W_3$ algebra, their eigenvalues under $L_0$ and $W_0$ being respectively
\begin{equation}
  \label{eq:conf_weights}
  \begin{split}
    \Delta_{\vect\alpha} &= 
    \frac{1}{2}\vect\alpha \cdot \left(\vect\alpha - 2 \vect\alpha_0 \right) \,, \\ 
    w_{\vect\alpha} &=
    \sqrt{\frac{12}{8 -15 \vect\alpha_0^2}}
    \ \prod_{i=1}^3 \left[ ( \vect\alpha -\vect\alpha_0) \cdot \vect h_i \right] \,.  \\
  \end{split}
\end{equation}
The (fully) degenerate field $\Phi_{\vect{n},\vect{m}}$ of the minimal model corresponds the following value of $\vect\alpha$
\begin{align}
  \label{eq:beta_def}
  \vect\alpha \begin{pmatrix} n_1 & m_1 \\ n_2 & m_2  \end{pmatrix} 
  &=  \vect{\alpha_0} - \alpha_+ \vect{n} - \alpha_- \vect{m} \nn \\
  & = \left[ (1-n_1)\alpha_+   + (1-m_1) \alpha_- \right] \vect\omega_1
  + \left[ (1-n_2)\alpha_+   + (1-m_2) \alpha_- \right] \vect\omega_2 \,.
\end{align}

While rational, the minimal models $W_3(p,q)$ are only unitary for $q = p+1$, and in this case they are equivalent to the following GKO coset\cite{petkova96} 
\begin{equation}
  W_3(p,p+1) = \frac{\wh{\mathfrak{su}}(3)_{k} \otimes \wh{\mathfrak{su}}(3)_1}{\wh{\mathfrak{su}}(3)_{k+1}}, \qquad p = k+3 \,.
\end{equation}

\subsection{Coulomb gas description of the FPL model}
\label{sec:CG}

The continuum limit  of loop models is rather well understood and typically yields a bosonic action coupled to a background curvature \cite{FODA1989643}. The scaling limit of the $A_2^{(1)}$ loop model in the case $n = \nt$ has been studied extensively in \cite{kondev1996operator}. Through a mapping to Coulomb gas, (part of) the spectrum and conformal dimensions were obtained.  We first present briefly these results, before extending them to the generic case $\nt \neq n$. When then focus on the particular value $\nt = n^2 -2$, for which an extended $W_3$ symmetry is expected. While the Coulomb gas obtained for the FPL loop model is not exactly equivalent to the $W_3$ Coulomb gas on generic Riemann surfaces, we argue that these two theories become equivalent on the flat cylinder. 

\subsubsection{The FPL model for $n  = \nt$}
\label{sec:CG-simple}

\begin{figure}
  \centering
  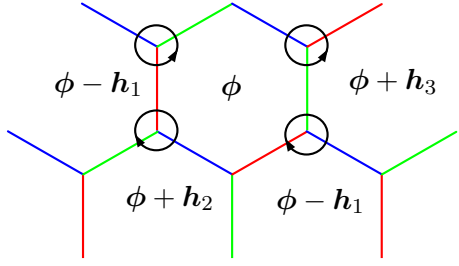
  \caption{Mapping from the three-color model to a height model.}
  \label{fig:loop_to_heights}
\end{figure}

Later in this section we will focus on the flat cylinder, but for now we consider a generic (i.e. non flat) Riemann surface of genus zero. As pointed out in \cite{FODA1989643}, this is extremely instructive in order to understand the coupling to curvature. The flat torus will be considered in \secref{torus}.  Before extending to $n < 2$, let us first consider the case $n = \nt =2$ on the hexagonal lattice, in which the partition function simply counts the the number of three-colourings (see \secref{loop_model}). Let $\vect\phi$ be a two-component discrete height field living on the dual lattice, for which the loops are level lines : $\vect\phi$ varies by $\pm 2\pi \vect{h}_i$, $i=1,2,3$ when crossing an edge of one of the three colors (the sign depends on the orientation of the edge, defined by the bi-partition of the lattice, see \figref{loop_to_heights}). In the scaling limit, this height field renormalizes towards a two-component boson with compactification lattice $\mathcal{R}$ \eqref{eq:R} and action 
\begin{equation}
  \label{eq:lagrangian_1}
  \begin{aligned}
    S_0 &= \frac{1}{8\pi} \int d^2x \,\sqrt{|\gamma|} \ \partial_{\mu} \vect{\phi} \cdot \partial^{\mu}\vect{\phi} \,, \\
    \vect\phi &\equiv \vect\phi + 2\pi\mathcal{R} \,,
  \end{aligned}
\end{equation}
where $\gamma$ is the metric of the underlying Riemann surface. 
The loop fugacity can be changed to $0 \leq n  \leq 2$ by introducing a coupling constant $g$ and a background vector charge
\begin{equation}
  \vect \beta_0=\frac{1-g}{2\sqrt{g}} \vect{\rho} \,.
\end{equation}
The action is modified to  
\begin{equation}
  \label{eq:lagrangian_2}
  \begin{aligned}
    S &= \frac{1}{8\pi} \int d^2x \, \sqrt{|\gamma|} \, \Big[
    g \times \partial_{\mu} \vect{\phi}\cdot \partial^{\mu}\vect{\phi}
    + 2 i R(x) \sqrt{g}\, \vect\beta_0\cdot \vect\phi(x)
    \Big] \,, \\
    \vect\phi &\equiv \vect\phi + 2\pi\mathcal{R} \,,
  \end{aligned}
\end{equation}
where $R(x)$ is the scalar curvature. The renormalized coupling constant $g$ is related to the weight of contractible loops, and is found to be 
\begin{equation}
  n = -2 \cos \pi g, \qquad g \in [0,1] \,.
\end{equation}
The coupling to the background curvature is required to obtain the same weights $n$ for all loops, and compensates for the deficiency angle of loops enclosing a non zero curvature \cite{FODA1989643}. On a genus zero surface the total curvature is $\int d^2x \,\sqrt{\gamma} \, R = 8\pi$.
The effect of the coupling to curvature is to modify the stress-energy tensor, lowering the central charge 
\begin{equation}
  T(z) = -\frac{g}{2} \normal{(\partial \vect{\varphi} ) ^2} 
  + i \sqrt{g}\vect\beta_0 \cdot \partial^2\vect{\varphi} \,,
  \qquad c = 2 - 12\, \vect\beta_0 \cdot \vect\beta_0 = 2 - \frac{6(1-g)^2}{g} \,,
\end{equation}
where we have used the convention $\vect\phi(z,\bar z) = \vect\varphi(z) + \bar{\vect\varphi}(\bar z)$ for the holomorphic and anti-holomorphic parts of the free field. For the partition function, charge neutrality requires the introduction of a neutralizing charge $2\sqrt{g}\vect\beta_0$ at some reference point:
\begin{equation}
  Z = \int [D\vect\phi] \ \exp\left[ -S(\vect\phi) + 2i\sqrt{g}\vect\beta_0 \vect\phi(x_0) \right] \,. 
\end{equation}

Let us specialize to the infinite flat cylinder, where the curvature is concentrated at $\pm \infty$. In this case, we choose $x_0=+\infty$. Then the action \eqref{eq:lagrangian_2} simplifies to
\begin{align}
  \label{eq:lagrangian_flat_cylinder}
  S_{\mathrm{cyl}} = \frac{1}{2\pi} \int d^2x \, \left\{ g\, \partial \vect{\phi}  \cdot \bar{\partial} \vect{\phi}  +i \sqrt{g}\vect\beta_0 \cdot \left[ \vect{\phi} (\infty) + \vect{\phi} (-\infty) \right] \right\}\,.
\end{align}
After combining the factors from the curvature terms and the neutralizing charge, the partition function will then correspond to the two-point function 
\begin{equation}
Z_{\rm cyl}(n=\nt)=\aver{e^{+i \sqrt{g}\vect\beta_0 \cdot \vect\phi(\infty)]} \times e^{-i \sqrt{g}\vect\beta_0 \cdot \vect\phi(-\infty)}}_0 \,,
\end{equation}
where $\aver{\dots}_0$ denotes the path integration with the simple bosonic action \eqref{eq:lagrangian_1}. Back to the lattice description, these two operators correspond to introducing an oriented ``seam line'' running along the axis of the cylinder, that assigns the correct weight to non-contractible loops.

Vertex operators in the theory~\eqref{eq:lagrangian_2} are defined as
\begin{equation}
  V_{\vect\alpha}= \normal{e^{i\sqrt{g} \vect\alpha \cdot \vect\phi}} \,,
\end{equation}
and they have conformal dimensions
\begin{equation}
  h_{\vect\alpha}=\bar h_{\vect\alpha}=\frac{1}{2}(\vect\alpha^2-2\vect\beta_0\cdot\vect\alpha) \,.
\end{equation}

\subsubsection{The FPL model for $n  \neq \nt$}
\label{sec:CG-eff}

On a surface of genus zero, all closed loops are in the same homotopy class, except if one marks two points on the surface, say $x_0$ and $y_0$. Then, those closed loops which split the points $x_0$ and $y_0$ become non-contractible, and we want to assign them a weight $\nt=2\cos\mu$. One may think of changing the amplitude of the curvature term in the action \eqref{eq:lagrangian_2}, but this would change the loop fugacities as soon as loops enclose a non-zero curvature, so this procedure is wrong for generic surfaces of genus zero.

Rather, we keep the Coulomb-Gas action \eqref{eq:lagrangian_2}, and insert the vertex operators $V_{\vect\beta_0-\vect\alpha_0}(y_0)$ and  $V_{\vect\beta_0+\vect\alpha_0}(x_0)$, with $\vect\alpha_0=\frac{\mu}{2\pi\sqrt{g}} \vect\rho$. Both of these operators have conformal dimensions $h=\bar h=(\vect\alpha_0^2-\vect\beta_0^2)/2$.

If we consider again the cylinder, and take $x_0=+\infty$ and $y_0=-\infty$, we obtain a theory with effective central charge:
\begin{equation} \label{eq:ceff1}
  c_{\mathrm{eff}} = c - 24h_{\vect\beta_0-\vect\alpha_0} = 2-12 \, \vect\alpha_0^2 = 2 - \frac{6(\mu/\pi)^2}{g} \,.
\end{equation}
In this theory, the effective conformal dimensions of $V_{\vect\alpha}$ are found by shifting the vertex charge by $(\vect\beta_0-\vect\alpha_0)$:
\begin{equation}
  \Delta_{\vect\alpha} = \bar\Delta_{\vect\alpha} = h_{\vect\alpha+\vect\beta_0-\vect\alpha_0}-h_{\vect\beta_0-\vect\alpha_0}
  = \frac{1}{2}(\vect\alpha^2 - 2\vect\alpha_0 \cdot \vect\alpha) \,.
\end{equation}
In the following, we shall restrict our study to the case prescribed by the Hecke algebra and the Markov trace (see \secref{Hecke}):
\begin{equation} \label{eq:fugacities}
  n=2\cos\lambda \,, \qquad \nt=2\cos 2\lambda \,, \qquad 0<\lambda<\pi \,.
\end{equation}
The effective Coulomb gas theory will then have:
\begin{equation} \label{eq:param}
  g=\frac{\pi-\lambda}{\pi} \,,
  \qquad \vect\alpha_0= (\alpha_+ + \alpha_-) \vect\rho \,,
  \qquad \alpha_-=\sqrt{g} \,, \qquad \alpha_+=-1/\sqrt{g} \,.
\end{equation}
The central charge in this case reads:
\begin{equation} \label{eq:ceff2}
  c_{\mathrm{eff}}=2-\frac{24(1-g)^2}{g} \,.
\end{equation}

\subsection{Operators and spectrum of the loop model}
\label{sec:spectrum}

The bulk spectrum of the FPL model with fugacities \eqref{eq:fugacities} can be described by defining the theory of \secref{CG-eff} on the cylinder, and considering separately the sectors of given ``magnetic charge'' $\vect q \in \mathcal{R}$, where $2\pi\vect q$ is the defect picked by the field $\vect\phi$ along the circumference of the cylinder. In this section, we describe the spectrum of primary operators.

\begin{itemize}

\item In the ``purely electric'' sector, i.e. the sector for $\vect q=0$, non-contractible loops may occur. The primary operators are all scalar vertex operators $V_{\vect\alpha}$ allowed by the compactification condition $\vect\phi \equiv \vect\phi + 2\pi \mathcal{R}$. For the vertex operator $V_{\vect\alpha}= \normal{e^{i\sqrt{g} \vect\alpha \cdot \vect\phi}}$ to be single-valued, one needs to impose $\vect\alpha \in \mathcal{R}^*/\sqrt{g}$. Comparing to \eqref{eq:beta_def} and \eqref{eq:param}, we end up with vertex charges of the form
\begin{equation} \label{eq:alpha1}
  \vect\alpha = (n_1\vect\omega_1+n_2\vect\omega_2)/\sqrt{g}
  = \vect\alpha \begin{pmatrix} 1+n_1 & 1 \\ 1+n_2 & 1  \end{pmatrix} \,,
  \qquad (n_1,n_2) \in \mathbb{Z}^2 \,.
\end{equation}
This ``purely electric'' sector, which includes the ground state, appears for lattices of width $N$ multiple of three. Due to the conservation laws, $N/3$ lines of empty edges propagate along the axis of the cylinder.

\item The sector where an additional $2k$ lines of empty edges propagate has a magnetic charge $\vect q=k(2\vect h_2-\vect h_1-\vect h_3)=k(\vect e_2-\vect e_1)$. Note that $(\vect e_2-\vect e_1)\mathbb{Z}$ is the set of vectors $\vect q \in \mathcal{R}$ such that $\vect q \cdot \vect\rho=0$. Like in the $\vect q=0$ sector, non-contractible loops are allowed. The magnetic charge $\vect q=k(\vect e_2-\vect e_1)$ may be combined with a vertex operator of charge $\vect\beta \in \mathcal{R}^*/\sqrt{g}$. The corresponding eigenvalues of $(L_0,\bar L_0, W_0, \bar W_0)$ may be computed from the Gaussian action by standard methods, and one gets
\begin{equation} \label{eq:Delta-w}
  \Delta = \Delta_{\vect\alpha} \,, \qquad \bar\Delta = \Delta_{\bar{\vect\alpha}} \,,
  \qquad w = w_{\vect\alpha} \,, \qquad \bar w = w_{\bar{\vect\alpha}} \,,
\end{equation}
where
\begin{equation} \label{eq:alpha2}
  \vect\alpha = \vect\alpha \begin{pmatrix} 1+n_1 & 1-\frac{3k}{2} \\ 1+n_2 & 1+\frac{3k}{2}  \end{pmatrix} \,,
  \qquad
  \bar{\vect\alpha} = \vect\alpha \begin{pmatrix} 1+n_1 & 1+\frac{3k}{2} \\ 1+n_2 & 1-\frac{3k}{2}  \end{pmatrix} \,,
  \qquad (n_1,n_2) \in \mathbb{Z}^2 \,.
\end{equation}
The conformal spin is
\begin{equation}
  s = \Delta_{\vect\alpha} - \Delta_{\bar{\vect\alpha}} = \frac{1}{2}(\vect\alpha+\bar{\vect\alpha}-2\vect\alpha_0) \cdot(\vect\alpha-\bar{\vect\alpha}) \,,
\end{equation}
which yields, for the above values of the vertex charges, $s=3k(n_1-n_2)$. Hence, this sector contains an infinity of scalar operators, including the most relevant one, obtained by setting $n_1=n_2=0$.

\item Any sector with one or more loop strands propagating has a magnetic charge $\vect q$ with $\vect q \cdot \vect\rho \neq 0$. For instance, the combination of $2\ell$ strands and $2k$ extra empty lines gives a magnetic charge $\vect q=\ell(\vect e_1+\vect e_2) + k (\vect e_2-\vect e_1)$. A generic charge $\vect q \in \mathcal{R}$ satisfying $\vect q \cdot \vect\rho \neq 0$ can be written
\begin{equation}
  \vect q=q_1 \vect e_1 + q_2 \vect e_2 \,,
  \qquad \text{where} \quad (q_1,q_2) \in \mathbb{Z}^2 \,,
  \qquad \text{and} \quad q_1+q_2 \neq 0 \,.
\end{equation}
In this case, because of the non-intersecting nature of loops, there are no non-contractible loops. Moreover, the oriented loop strands may wind around the cylinder, which produces unwanted factors due to the vertex charges at $\pm\infty$. Hence, the magnetic defects in this sectors must be combined with ``electric charges'' $\pm \vect\alpha_0$ to compensate this effect. One ends up with eigenvalues~\eqref{eq:Delta-w}, parametrised by the charges:
\begin{equation} \label{eq:alpha3}
  \vect\alpha = \vect\alpha \begin{pmatrix} n_1 & q_1-\frac{q_2}{2} \\ n_2 &  q_2-\frac{q_1}{2} \end{pmatrix} \,,
  \qquad
  \bar{\vect\alpha} = \vect\alpha \begin{pmatrix} n_1 & -q_1+\frac{q_2}{2} \\ n_2 & -q_2+\frac{q_1}{2}  \end{pmatrix} \,,
  \qquad (n_1 \vect\omega_1 + n_2 \vect\omega_2) \in \mathcal{R}_{\vect q}^* \,.
\end{equation}
In this expression, $\mathcal{R}_{\vect q}=\mathbb{Z} \vect q + \mathbb{Z}(\vect e_1-\vect e_2)$ is the lattice of allowed defects along the axis of the cylinder, and $\mathcal{R}_{\vect q}^*$ is its reciprocal lattice:
\begin{equation}
  \mathcal{R}_{\vect q}^* = \left\{
    n_1 \vect\omega_1 + n_2 \vect\omega_2 \,,
    \quad (n_1,n_2) \in \left( \frac{\mathbb{Z}}{q_1+q_2} \right)^2 \,,
    \quad (n_1+n_2) \in \mathbb{Z}
  \right\} \,.
  \label{eq:Rq}
\end{equation}
This includes operators with rational, non-integer Kac indices $n_1$ and $n_2$ in \eqref{eq:alpha3}. However, the conformal spin remains an integer:
\begin{equation}
  s = -(n_1+n_2)(q_1+q_2) - (n_1-n_2)(q_1-q_2) \,.
\end{equation}

\end{itemize}

\subsection{Time-like Toda field theory}

The imaginary affine Toda field theory \cite{fateev07} is defined by the action
\begin{equation}
  \label{eq:toda_action}
    S = \frac{1}{8\pi} \int d^2x \, \sqrt{|\gamma|} \, \Big[
    g \times \partial_{\mu} \vect{\phi}\cdot \partial^{\mu}\vect{\phi}
    + 2 i R(x) \sqrt{g}\, \vect\alpha_0\cdot \vect\phi(x)
    + \kappa \left(\!\normal{e^{-i\vect\phi \cdot \vect{e}_1}} + \normal{e^{-i\vect\phi \cdot \vect{e}_2}} \right)
    \Big] \,,
\end{equation}
where the conventions~\eqref{eq:param} have been used, and $\kappa$ is an arbitrary scale parameter. It is conformal, with $W_3$ symmetry, and central charge $c = 2 - 12 \vect\alpha_0^2$. The main difference with the Coulomb-Gas theory \eqref{eq:lagrangian_2} is that the field $\vect\phi$ is \textit{not} compactified. As a consequence, no magnetic defect is allowed, and the spectrum of primary operators is continuous, consisting in all vertex operators
\begin{equation}
  V_{\vect\alpha} = \normal{e^{i\sqrt{g}\vect\alpha \cdot \vect\phi}} \quad,
  \qquad \vect\alpha \in \mathbb{R}^2 \,,
\end{equation}
 with conformal dimensions
\begin{equation}
  \Delta_{\vect\alpha} = \bar\Delta_{\vect\alpha} = \frac{1}{2}(\vect\alpha^2 - 2\vect\alpha_0 \cdot \vect\alpha) \,.
\end{equation}
Note that the vertex operators in the action~\eqref{eq:toda_action} both have dimensions $\Delta=\bar\Delta=1$. We conjecture that the time like Toda theory describes the two- and three-point functions of generic ``purely electric'' operators which change the loop fugacities in the FPL model, in the same way as the time-like Liouville theory does for the O($n$) model.

\section{Numerical study}
\label{sec:numerics}

\subsection{Loop model Hamiltonian}


In order to check the previous results we use exact diagonalisation methods on systems of sizes $N=6$ to $N=18$. For simplicity, we restrict ourselves to the case when $N$ is multiple of 3. We consider the Hamiltonian of the loop model transfer matrix with periodic boundary conditions:
\begin{equation}
  H_N \propto \left.\frac{d \log t_N(u)}{du}\right|_{u=0}
  = -\sum_{j=1}^N U_j \,,
\end{equation}
where $U_j$ is given by~\eqref{eq:Uj-loop}, and we restrict specifically to the case $\nt=n^2-2$. The Hilbert space for the Hamiltonian is generated by non-intersecting link patterns which allow vacancies: see next section.

In the scaling limit, one expects from conformal invariance the following form of the energy and momentum:
\begin{align}
  E_N &\sim N e_{\mathrm{bulk}} + \frac{2\pi v_f}{N} \left(\Delta + \bar\Delta-\frac{c}{12} \right) \,,
  \label{eq:E} \\
  P_N &\sim \mathrm{const} + \frac{2\pi}{N} \left(\Delta - \bar\Delta \right) \,,
  \label{eq:P}
\end{align}
where $e_{\mathrm{bulk}}$ is the non-universal bulk energy density.
The Fermi velocity $v_f$ in~\eqref{eq:E} may be inferred from the expression of the embedding angle $\theta=\frac{2}{3}\pi u/\lambda$ (see \secref{loop_model}: 
\begin{equation}
  v_f = \frac{2 \pi \sin\lambda}{3 \lambda} \,.
\end{equation}

\subsection{Loop model Hilbert space}

By analogy with the standard modules in the representation theory of the (periodic) Temperley-Lieb algebra, we consider the representation of the Hamiltonian $H_N$ on vector spaces generated by link patterns. In this work, we are not treating mathematically the representation theory of the periodic version of the $\mathfrak{sl}(3)$ Hecke algebra. Rather, we choose empirically a family of representations (which we call {\it loop sectors}), and show numerically that the associated conformal weights match the predictions from the Coulomb Gas approach.

\begin{figure}
  \centering
  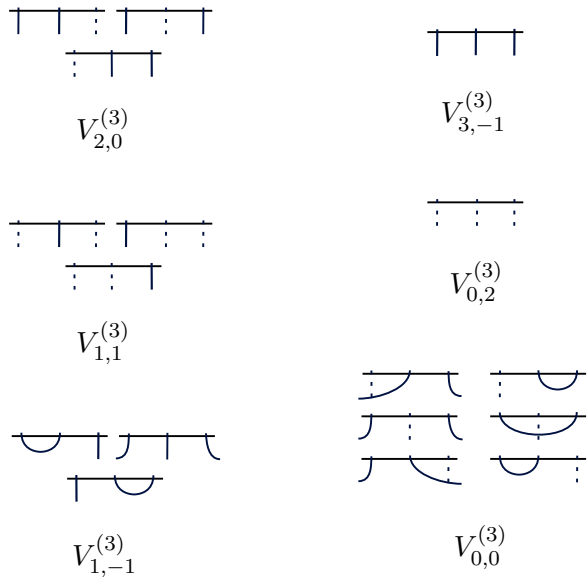
  \caption{The sectors $V_{d,v}^{(N)}$ of the loop model for $N=3$.}
  \label{fig:sectors}
\end{figure}

We fix the system size $N$, multiple of 3. For any $d\in\{0,1,\dots N\}$ and $v\in\{-\frac{N}{3}, -\frac{N}{3}+1, \dots \frac{2N}{3}-d\}$ such that $d \equiv v \mod 2$, let $V_{d,v}^{(N)}$ be the vector space generated by all link patterns with $d$ strands connected to infinity, and $(N/3+v)$ vacant sites: see \figref{sectors} The action of the generators~$U_j$ on $V_{d,v}^{(N)}$ is analogous to that of the periodic Temperley-Lieb on its standard modules: each plaquette of \eqref{eq:Uj-loop} (considered as acting from SW to NE), evolves the link pattern according to the graphical prescription, and introduces a factor $n$ or $\nt$ for every closed contractible or non-contractible loop. Note that two strands connected at infinity cannot get contracted under the action of the $U_j$'s.

In the continuum limit, i.e. in the two-component boson theory of~\secref{CG-eff}, we expect the low-energy part of $V_{d,v}^{(N)}$ to be described by the sectors of magnetic charges $\vect q$ which correspond to $d$ propagating strands and $N/3+v$ vacancies. For instance, for $d$ and $v$ even, the lowest-energy state of $V_{d,v}^{(N)}$ has magnetic charge $\vect q=\frac{1}{2}[d(\vect e_1+\vect e_2) + v(\vect e_2-\vect e_1)]$ (see \secref{spectrum}). The ground state is located in the sector $V_{0,0}^{(N)}$.

The constant contribution to the momentum in~\eqref{eq:P}, although it is not universal, reflects the three-fold structure of the lattice model. By numerical observation, we find that this contribution is determined by the Kac indices $(n_1,n_2)$ parametrising $(\Delta,\bar\Delta,w,\bar{w})$ in~\eqref{eq:alpha1}, \eqref{eq:alpha2} and \eqref{eq:alpha3}. It is zero if $(n_1-n_2)$ is a multiple of 3, and $\pm\frac{2\pi}{3}$ otherwise. Descendants of a primary field have the same constant contribution to the momentum as the primary field.
 
 


\subsection{Numerical results}

We compute the energies of $H_N$ in various subsectors of fixed momentum within the sectors $V_{d,v}^{(N)}$, and extract the scaling dimensions $(\Delta+\bar\Delta)$ using \eqref{eq:E}. We extrapolate the data from system sizes $L=6 \dots 18$ using Shank's method (see \cite{domb1983phase}). The sizes obtainable are limited by the memory needed by the Hilbert space (of size $\sim 10^6$ for $L=18$). The noticeable results we have obtained are:

\begin{itemize}

\item The expression~\eqref{eq:ceff2} for the central charge is confirmed.

\item The predictions from \secref{spectrum} on the scaling dimensions of primary operators in various sectors is confirmed. Fractional Kac indices, as predicted by \ref{eq:Rq} appear for non-zero magnetic charges. Contrary to the O($n$) case, they can appear even in sectors with zero conformal spin.

\item In addition to Virasoro descendants, we identify additional states, whose conformal dimensions match with descendants under $W_{n<0}$ modes.

\item Unlike in the O($n$) model, some electro-magnetic excitations [see \eqref{eq:alpha2}] can have vanishing conformal spin, even though their electric and magnetic charges are nonzero.

\item The finite-size effects can be quite important especially near $c=2$, due to the increased presence of logarithmic corrections, or when different states cross.

\item Similarly to the O($n$) model \cite{EstienneIkhlef15}, the descendants $L_{-1}\ket 0$ and $W_{-1}\ket 0$, although they have zero norm, do appear in the spectrum of the lattice Hamiltonian $H_N$. This suggests the existence of logarithmic CFT features of the FPL model.

\end{itemize}

\vspace{0.5cm}
\begin{minipage}{\linewidth}
  \centering
  \includegraphics[width=0.9\linewidth]{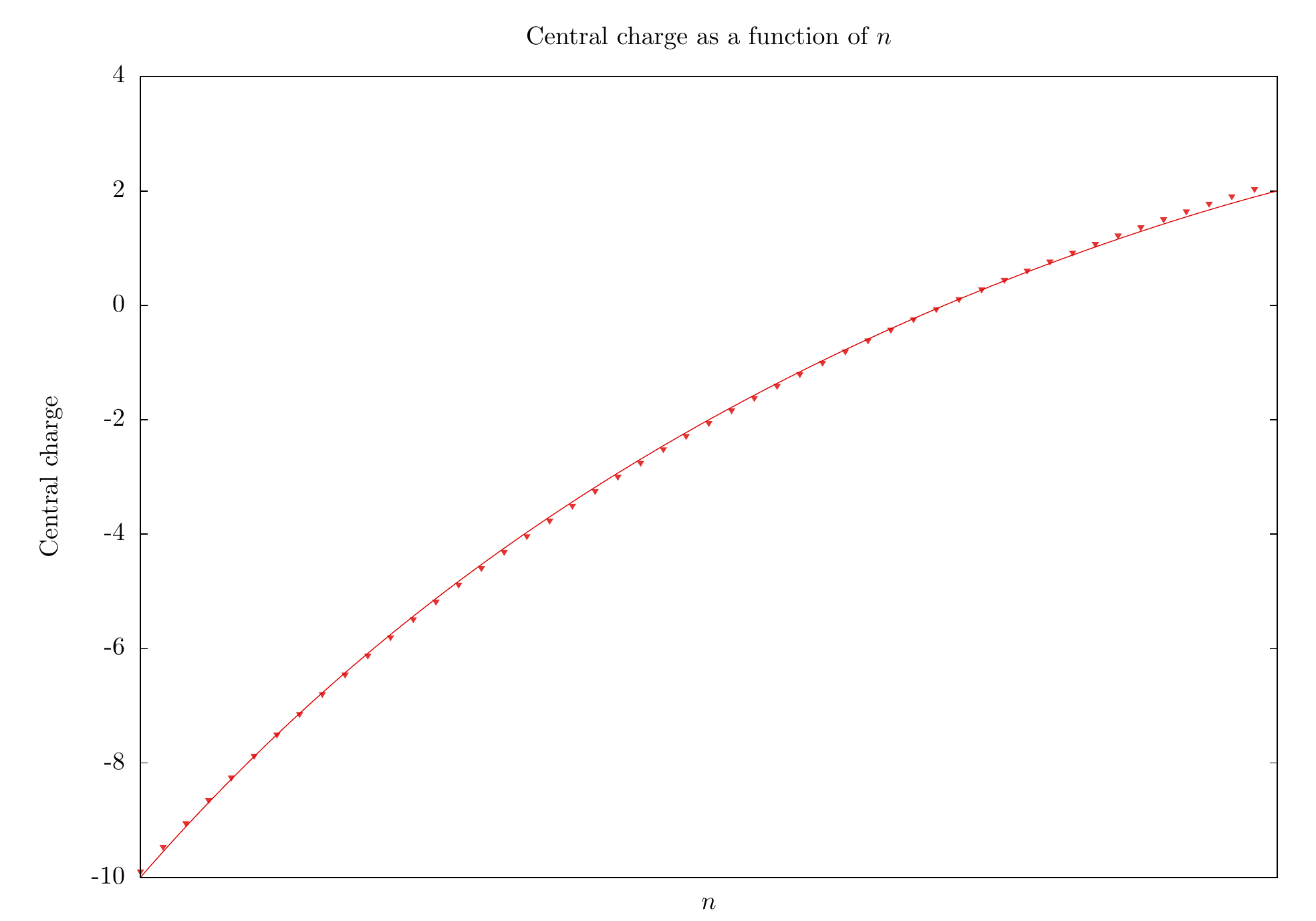}
  \captionof{figure}{Central charge of the FPL model for $\nt=n^2-2$, as a function of $n$. The numerical data for sizes  $N=6,9,12,15,18$ are compared to the analytic prediction~\eqref{eq:ceff2}.}
  \label{fig:central_charge}
\end{minipage}
\vspace{0.5cm}

\begin{minipage}{\linewidth}
  \centering
  \includegraphics[width=0.9\linewidth]{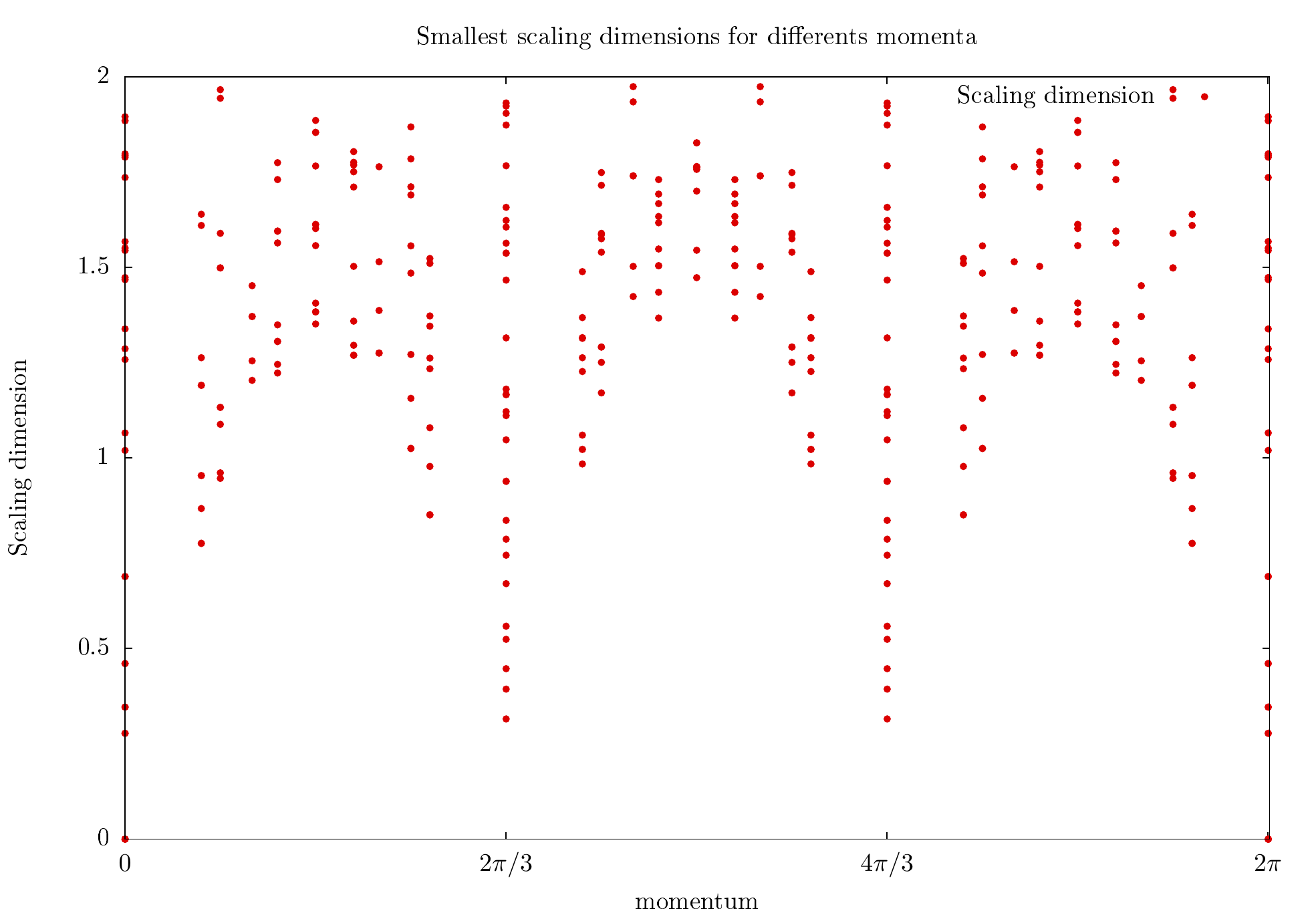}
  \captionof{figure}{The smallest scaling dimensions of the Hamiltonian as a function of the momentum, for $n = 1.7$. Multiple size are represented, for $N=6,9,12,15$ }
  \label{fig:weight_vs_impulsion}
\end{minipage}
\vspace{1cm}

\begin{minipage}{\linewidth}
  \centering
  \includegraphics[width=0.9\linewidth]{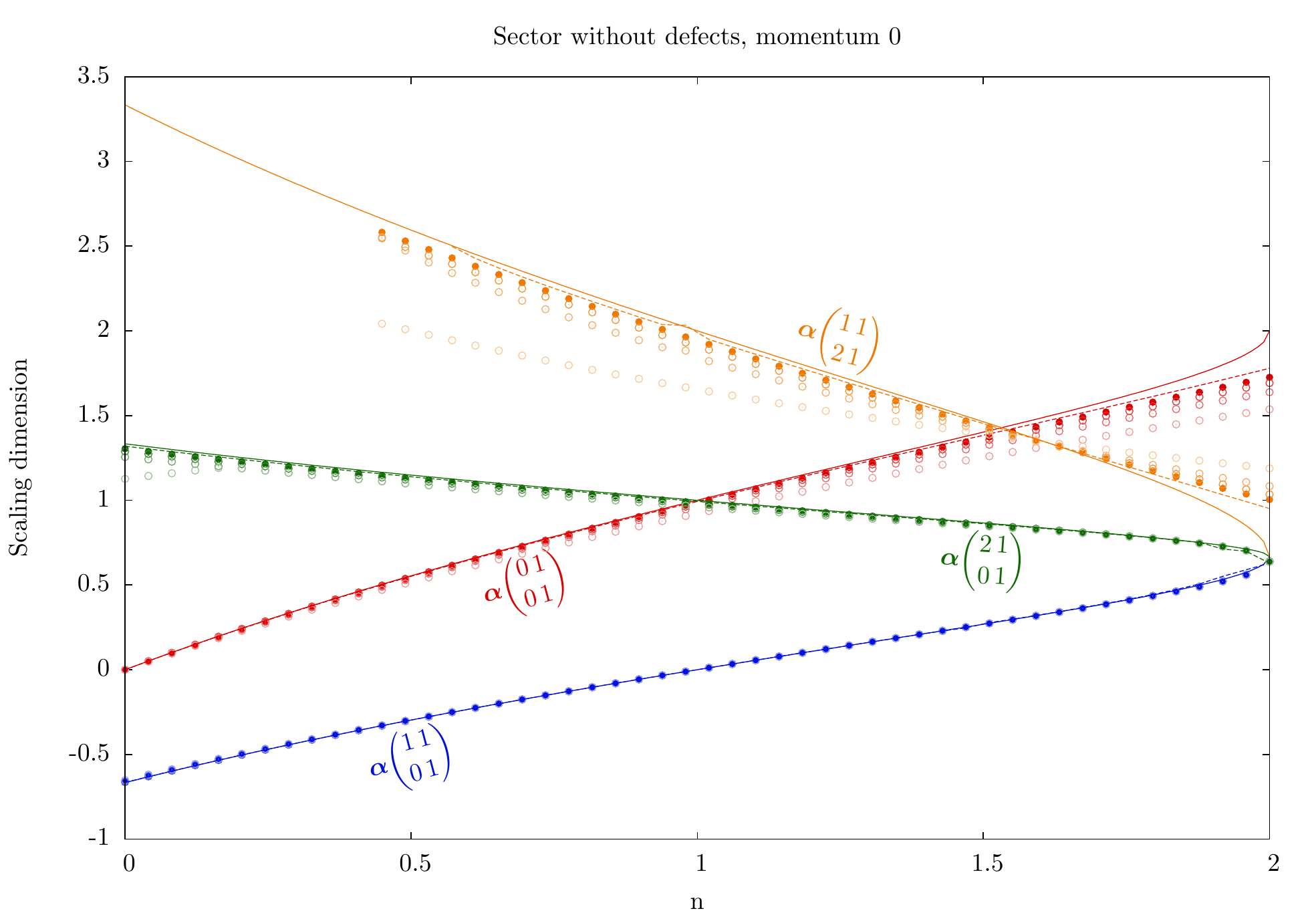}
  \captionof{figure}{The purely electric sector, $V^{(N)}_{0,0}$, momentum $P=0$. Dots are numerical results, darker dots corresponds to larger size, from $6$ to $15$, dotted lines their Shank's transform and the full lines represent the theory. The labels on the curve represent vertex charges and correspond to \eqref{eq:alpha1}.}
  \label{fig:electric_imp0}
\end{minipage}

\begin{minipage}{\linewidth}
  \centering
  \includegraphics[width=0.9\linewidth]{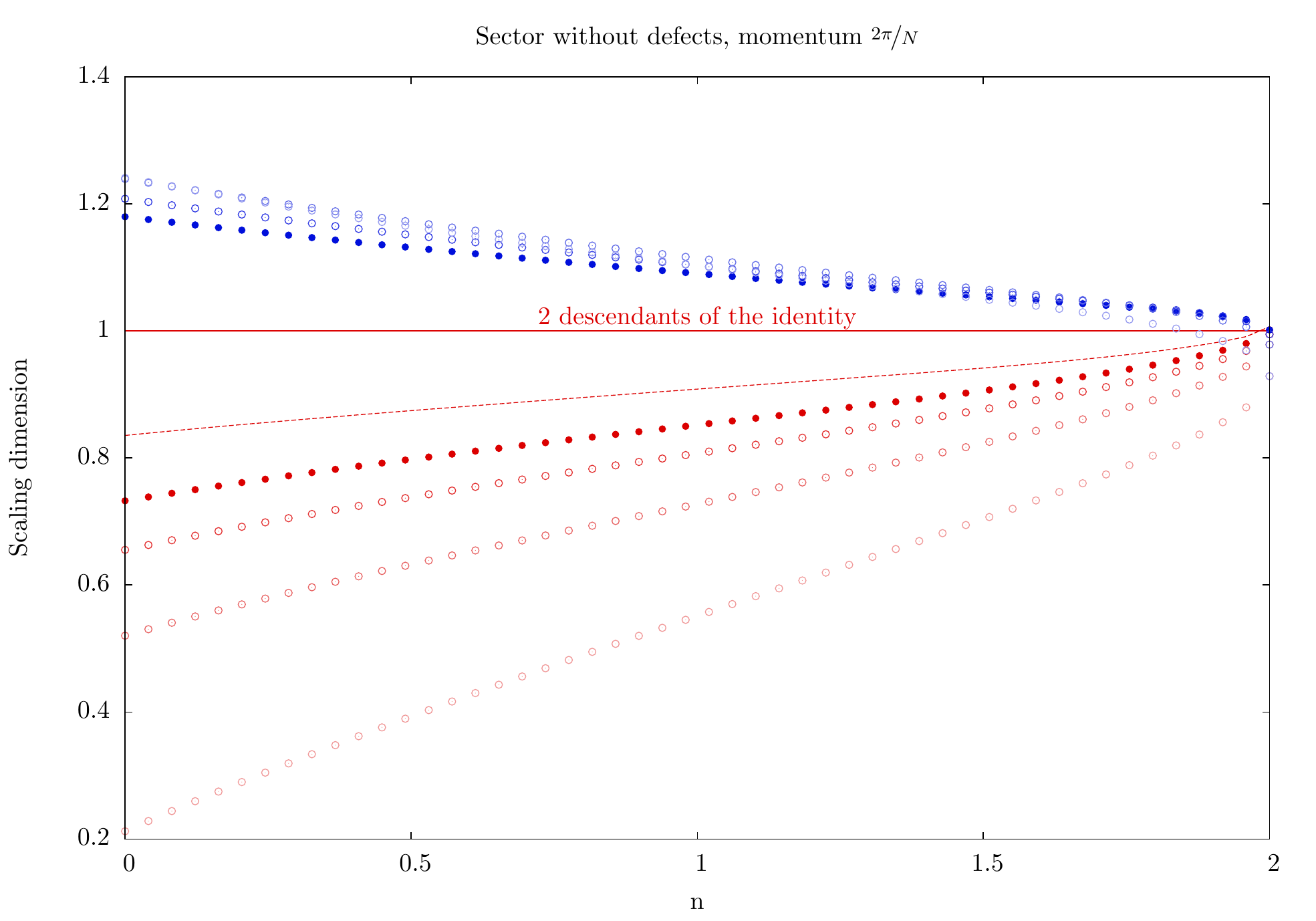}
  \captionof{figure}{The purely electric sector, $V^{(N)}_{0,0}$, momentum $P=\nicefrac{2\pi}{N}$. Dots are numerical results (lighter $N=6$ up to darker $N=15$ in step of $3$), dotted lines their Shank's transform and the full lines give the theory. While the convergence for descendant states is pretty bad, their position in the spectrum and their momentum label those two states as $L_{-1}\ket{0}$ and $W_{-1}\ket{0}$.}
  \label{fig:electric_imp1}
\end{minipage}

\begin{minipage}{\linewidth}
  \centering
  \includegraphics[width=0.9\linewidth]{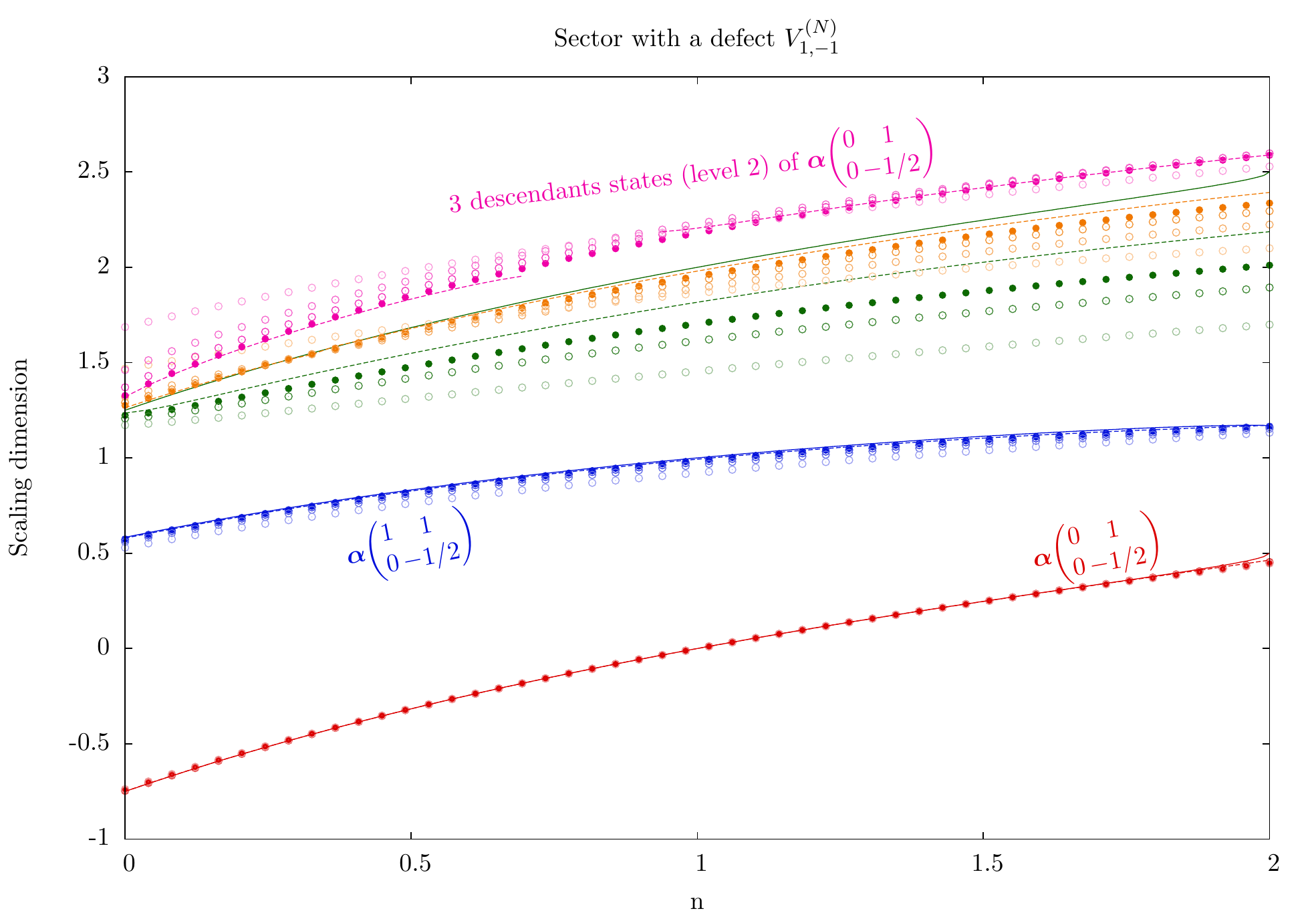}
  \captionof{figure}{Sector $V^{(N)}_{1,-1}$, momenta $P=0$ and $P=\nicefrac{2\pi}{3}$, same conventions as in \figref{electric_imp0} . The label on the curve have to be understood as the holomorphic part of a defect plus an electric charge, corresponding to the expression \eqref{eq:alpha3} with $q_1=1$ and $q_2=0$. As per \eqref{eq:Rq}, no fractional electric charge appears in this sector.}
  \label{fig:magnetic_B11}
\end{minipage}

\begin{minipage}{\linewidth}
  \includegraphics[width=0.9\linewidth]{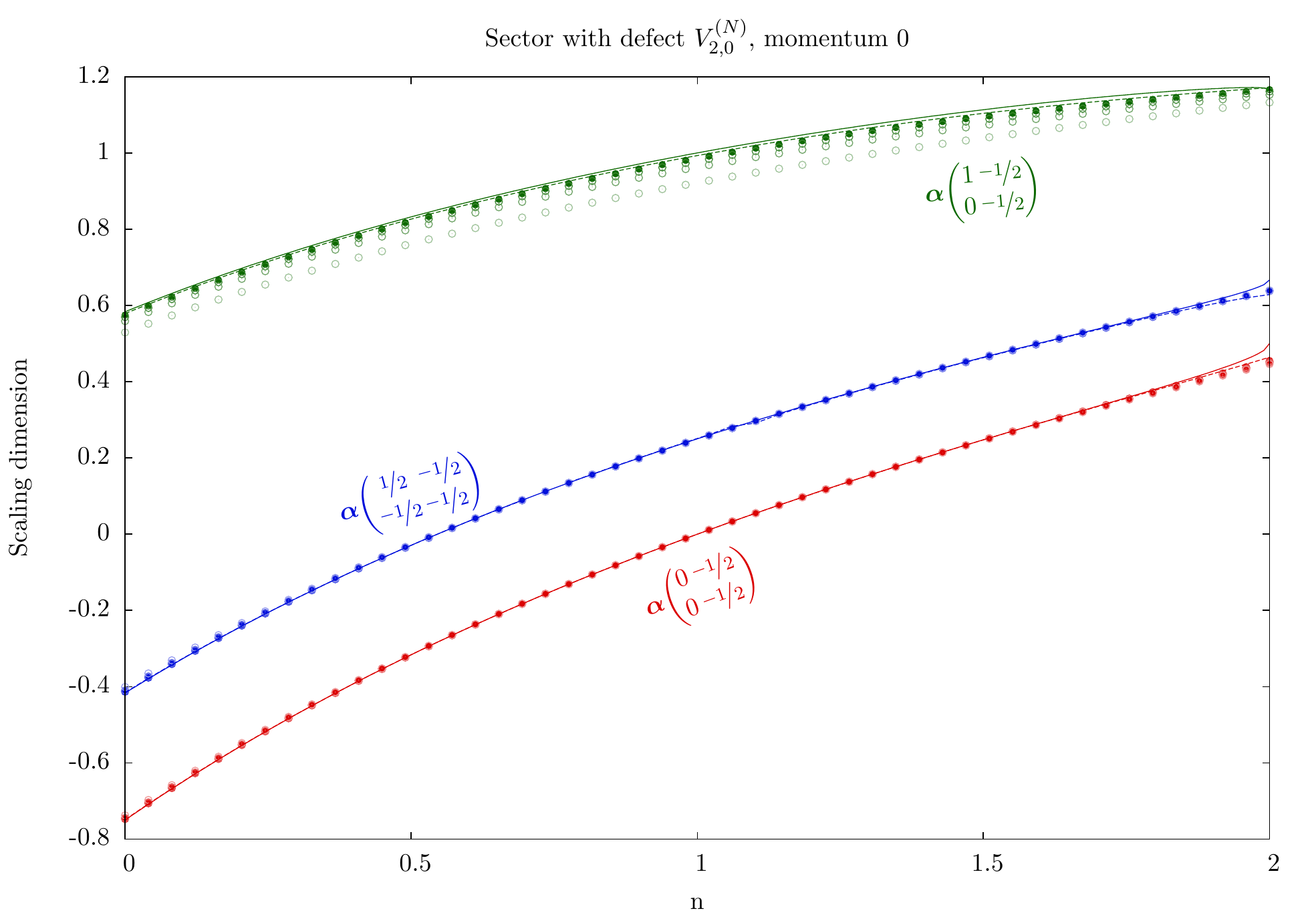}
  \captionof{figure}{Sector $V^{(N)}_{2,0}$ (two strands) momentum $0$, same conventions as \figref{electric_imp0}. Again the label have to be understood through \ref{eq:alpha3} with $q_1=q_2=1$. This allows for half-integer Kac indices $n_1$ and $n_2$.}
\end{minipage}

\begin{minipage}{\linewidth}
  \includegraphics[width=0.9\linewidth]{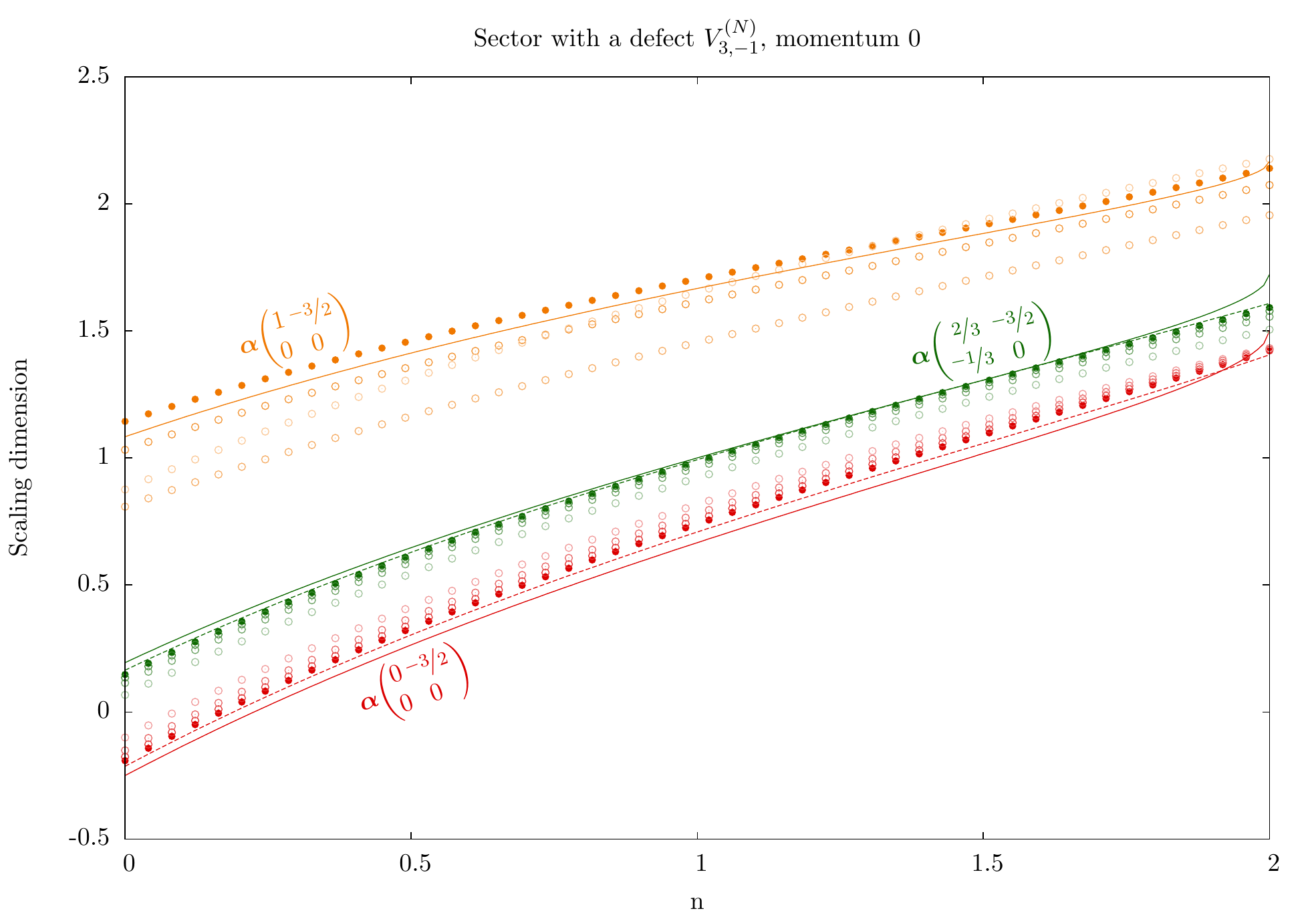}
  \captionof{figure}{Sector $V^{(N)}_{3,-1}$ (three strands added and one hole removed), momentum $0$, same conventions as \figref{electric_imp0}. The labels correspond to \eqref{eq:alpha3} with $q_1=2$ and $q_2=1$. This time electric charges live in $\mathbb{Z}/3$.}
  \label{fig:magnetic_three}
\end{minipage}

\begin{minipage}{\linewidth}
  \includegraphics[width=0.9\linewidth]{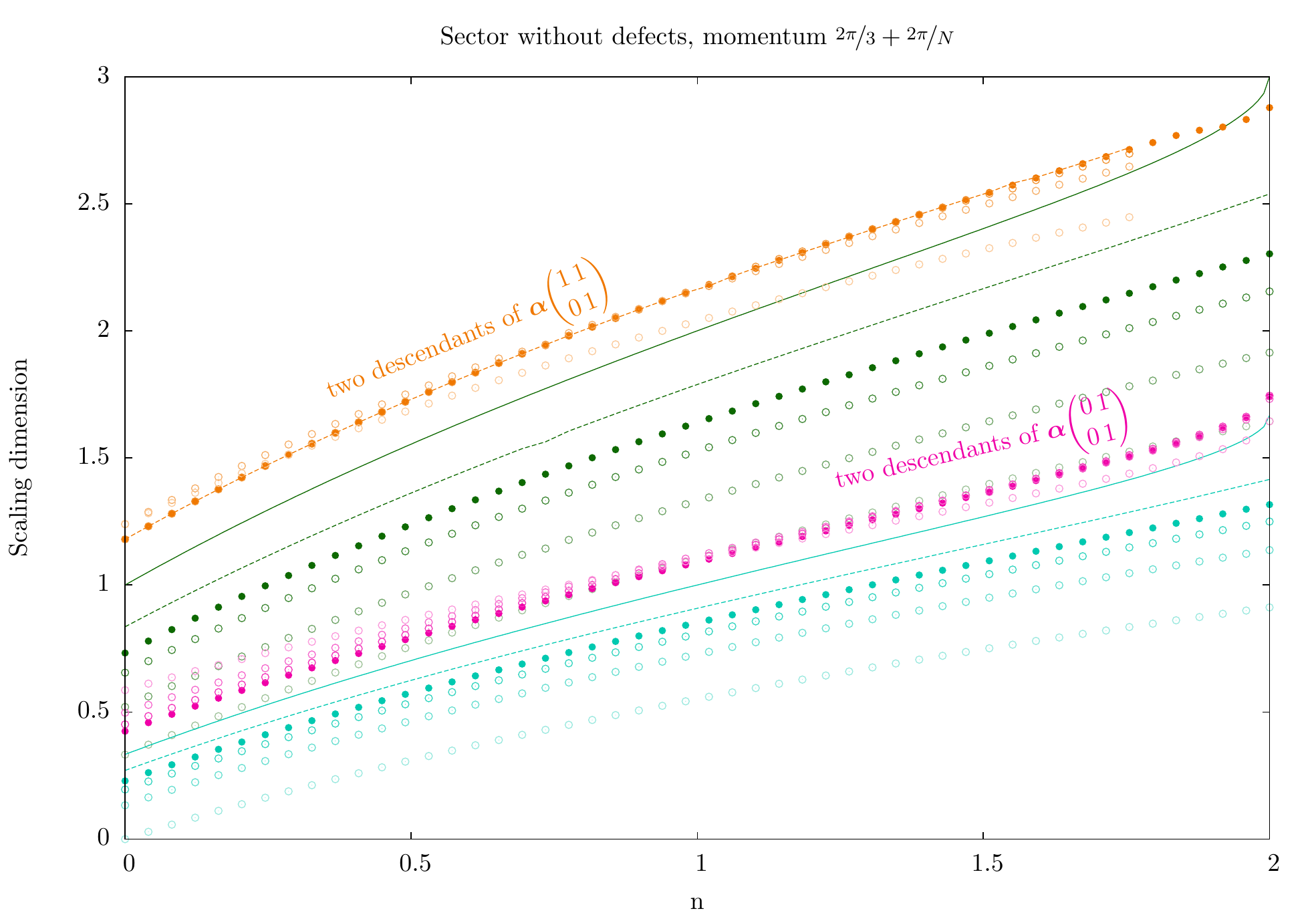}
  \captionof{figure}{The purely electric sector, $V^{(N)}_{0,0}$, momentum $P=\nicefrac{2\pi}{3} +  \nicefrac{2\pi}{N}$, same conventions as \figref{electric_imp0}. Again, descendants states tend to converge badly, but their degeneracy is consistent with the theory.}
\end{minipage}

\begin{minipage}{\linewidth}
  \includegraphics[width=0.9\linewidth]{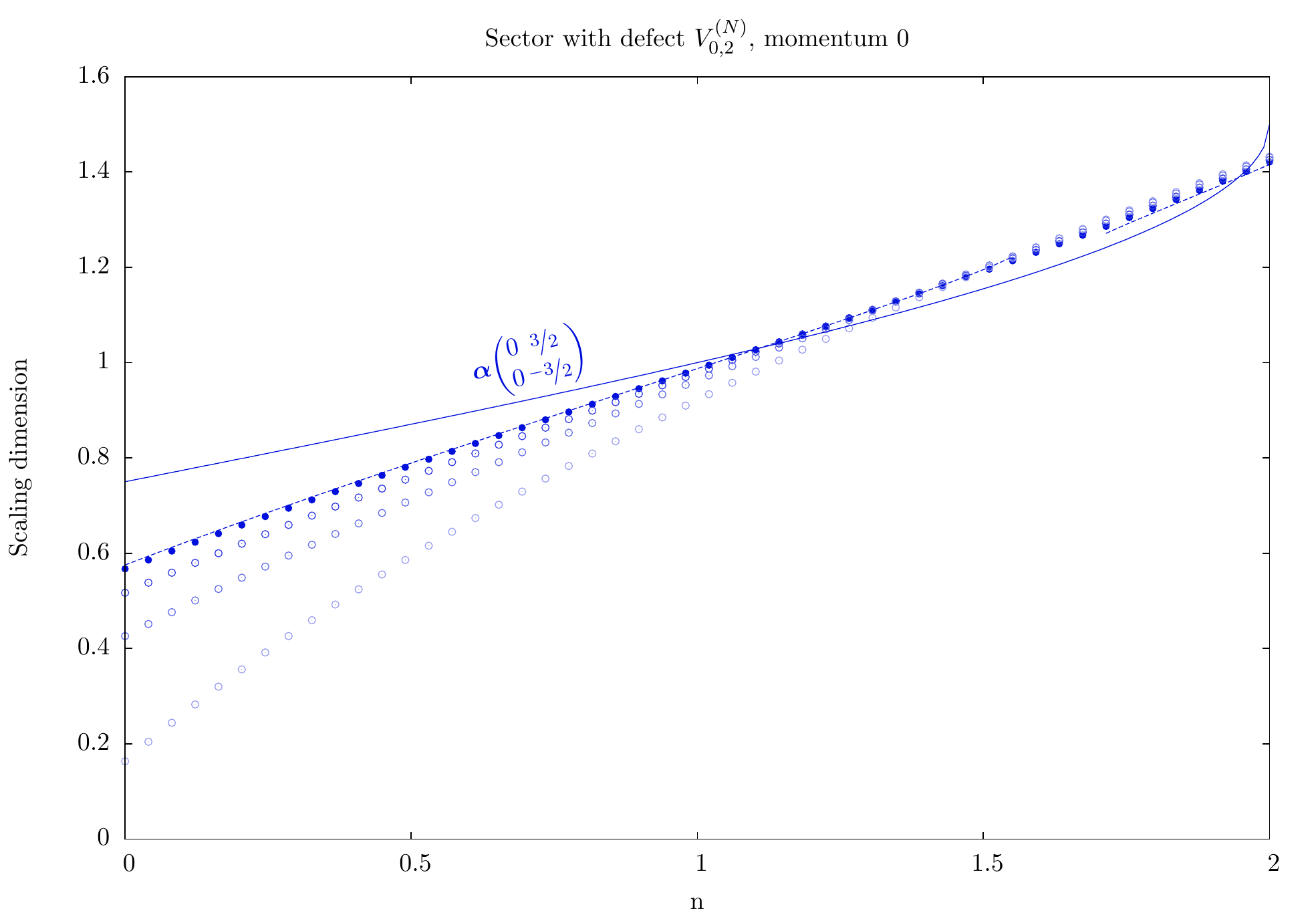}
  \captionof{figure}{Two hole sector $V^{(N)}_{0,2}$, same conventions as \figref{electric_imp0}. It is a magnetic sector with charge $\vect q$ perpendicular to $\vect{\rho}$, hence the vertex charges correspond to the case \eqref{eq:alpha2}. In the case $n=\nt=2$, the model enjoys an additional $\mathbb{Z}_3$ symmetry (cyclic permutations of the colors -- see \figref{loop_to_heights}), and the sectors $V^{(N)}_{0,2}$ and $V^{(N)}_{3,-1}$ become isomorphic.}
\end{minipage}

\section{The fully packed loop model on the torus}
\label{sec:torus}

On the torus, the partition function depends on every operators appearing in the theory. In this sense it's an important object to compute in order to understand the complete system. As usual we parametrise the torus by a complex number $\tau$, with $\operatorname{Im}(\tau) > 0$ such that $\mathbb{T} = \mathbb{C}/\left(\tau \mathbb{Z} + \mathbb{Z}\right)$. The partition function will be expressed as a function of $q=e^{2\compi \pi \tau}$.

\subsection{Partition function of the compact boson}

The partition function of the loop model can be obtained starting from the partition function of a compact boson, following \cite{difrancesco_saleur_zuber_coulomb}.
If $\vect\phi$ is a boson compactified on a given lattice $2\pi \mathcal{R}$, a configuration of the field will verify:
\begin{equation}
  \label{eq:compact_boson}
  \vect{\phi}(z + 1,\bar{z}+1) =   \vect{\phi}(z,\bar{z}) +  2\pi  \,  \vect{q} \,,
  \qquad \vect{\phi}(z+\tau,\bar{z} + \bar{\tau}) = \vect{\phi}(z,\bar{z}) +   2\pi \, \vect{q}' \,,
\end{equation}
where $\vect{q}, \vect{q}' \in \mathcal{R}^2$. The calculation of the torus partition function in the sector $(\vect{q},\vect{q}')$ is a standard exercise. Separating the quantum and classical degrees of freedom of $\vect{\phi}$, one finds 
\begin{equation}
  \label{eq:partition_function_b1_b2}
  Z_{\vect{q},\vect{q}'}(g,\tau) =  \frac{g}{2\operatorname{Im} \tau |\eta(\tau)|^4} \exp\left(-\frac{\pi g}{2 \operatorname{Im} \tau} \left|\vect{q} \tau - \vect{q}' \right|^2 \right) \,.
\end{equation}
The full partition function is expected to be of the form
\begin{equation}
  \label{eq:Z_general_torus}
  Z(\tau) = \sum_{(\vect{q},\vect{q}') \in \mathcal{R}^2} c_{\vect{q},\vect{q}'} Z_{\vect{q},\vect{q}'}(g,\tau) \,,
\end{equation}
where the $c_{\vect{q},\vect{q}'}$ are some constant coefficients.
For the compact boson theory~\eqref{eq:lagrangian_1}, all the sectors contribute with the same weight, and by convention $c_{\vect{q},\vect{q}'}=\text{Vol}(\mathcal{R})$, where $\text{Vol}(\mathcal{R}) = \sqrt{3}$ is the area of the unit cell of $\mathcal{R}$:
\begin{equation}
  Z_{c}[g, \mathcal{R}, \tau] =  \text{Vol}(\mathcal{R}) \sum_{(\vect{q},\vect{q}') \in \mathcal{R}^2} Z_{\vect{q},\vect{q}'}(g,\tau) \,.
\end{equation}
The global normalization has been chosen to recover the usual behavior $Z \sim (q \bar{q})^{-c/24}$ as $q \to 0$ of the full partition function. Indeed, using a Poisson summation to express the result in term of magnetic and electric charges yields
\begin{equation}
  \label{eq:Z_compact_boson_torus}
  Z_{c}[ g,\mathcal{R},\tau] = \frac{1}{|\eta(\tau)|^4}  \sum_{\vect{e} \in \mathcal{R}^*} \sum_{\vect{q} \in \mathcal{R}}q^{\delta(\vect{e},\vect{q})} \bar{q}^{\bar{\delta}(\vect{e},\vect{q})} \,.
\end{equation}
By analogy with the $\mathfrak{su}(2)$ case, we will call this model of
compact boson the $f$ model \cite{pasquier1987}. The  weights $\delta,\bar\delta$ correspond to the spectrum of the free compact boson:
\begin{equation}
  \label{eq:weight_torus}
  \delta(\vect{e}, \vect{q}) = \frac{1}{2}\left( \frac{\vect{e}}{\sqrt{g}} - \frac{1}{2}\sqrt{g} \vect{q} \right)^2
  \qquad \overline{\delta}(\vect{e}, \vect{q}) = \frac{1}{2}\left( \frac{\vect{e}}{\sqrt{g}} + \frac{1}{2}\sqrt{g} \vect{q} \right)^2 \,.
\end{equation}
We also recall the duality relation
\begin{align}
  Z_c[g, \mathcal{R}, \tau] =  Z_c[4/g, \mathcal{R}^*, \tau] \,.
\end{align}

\subsection{Partition function of the loop model}
\label{sec:partition_function_loop_torus}

In the loop model the coefficients $c_{\vect{q},\vect{q}'}$ in \eqref{eq:Z_general_torus} depend on the weight of the non-contractible loops, as was first established in the case of the $O(n)$ model \cite{difrancesco_saleur_zuber_coulomb}. The same kind of arguments apply to the FPL model. Let $\nt = 2\cos 2\pi e_0$ be the weight of non-contractible loops, then the partition function on the torus is
\begin{align}
  Z_{\textrm{loop}}\left[ g,e_0,\mathcal{R}\right]  =  \text{Vol}(\mathcal{R}) \times
  \sum_{(\vect{q},\vect{q}') \in \mathcal{R}^2}  \exp \left[  2i\pi e_0  (\vect{q}\cdot \vect{\rho}) \wedge (\vect{q}'\cdot \vect{\rho}  ) \right] \, Z_{\vect{q},\vect{q}'}(g)
\end{align}
where $a \wedge b$ stands for the greatest common divisor (gcd) of $a$ and $b$ (note that for any $\vect{q} \in \mathcal{R}$, one has $\vect{q}\cdot\vect\rho \in \mathbb{Z}$). The spectrum can be extracted using Poisson summation and a careful treatment of the $\gcd$ factor, as was done for the $O(n)$ model \cite{difrancesco_saleur_zuber_coulomb,read_saleur_sigma_01}. The detailed of this derivation for the FPL model can be found in Appendix \ref{ressumation}. The partition function takes the form
\begin{align}
  Z_{\textrm{loop}}[g, e_0, \mathcal{R}] 
  = \frac{(q\bar q)^{(2-c)/24}}{|\eta(\tau)|^4}  
    & \left[\sum_{\substack{ \vect{q} \in \mathcal{R} \\ m = 0 } }  \sum_{\vect{e} \in \mathcal{R}^*} q^{\Delta(\vect{e}- e_0 \vect{\rho},\vect{q})} \bar{q}^{\bar{\Delta}(\vect{e}-e_0 \vect{\rho},\vect{q})} \right.  \nn \\ 
  & \quad + \left. \sum_{\substack{ \vect{q} \in \mathcal{R} \\ m \neq 0 } } \sum_{k | m} \sum_{\vect{e} \in \wt{\mathcal{R}}_k^*} \Lambda(k,m,e_0) q^{\Delta(\vect{e},\vect{q})} \bar{q}^{\bar{\Delta}(\vect{e},\vect{q})} \right] \,,
  \label{loop-partition-function}
\end{align}
where we have introduced $m = \vect{q}\cdot\vect{\rho}$ for compactness. The lattice $\wt{\mathcal{R}}^*_k$ is defined as
\begin{equation}
  \wt{\mathcal{R}}^*_k = \left\{  a \vect{\omega}_1 + \frac{b\vect{\rho}}{k} \,,
    \quad (a,b) \in \mathbb{Z}^2 \,,
    \quad b \wedge k= 1
  \right\} \,,
\end{equation}
and the central charge and the exponents are
\begin{equation}
  \begin{aligned}
  \Delta(\vect e, \vect q) &= \delta(\vect e, \vect q) - \delta(\vect e_0,0) \,, \\
  \bar \Delta(\vect e, \vect q) &= \bar \delta(\vect e, \vect q) - \bar \delta(\vect e_0,0) \,, \\
  c &= 2-24 \Delta(\vect e_0,0) \,.
  \end{aligned}
\end{equation}
The function $\Lambda(k,m,e_0)$ can be found in Appendix \ref{ressumation}.

The expression~\eqref{loop-partition-function} confirms the results of \secref{spectrum}. Indeed, for a generic $W_3$ module of weights $(\Delta,w)$, the character reads
\begin{equation}
  \mathrm{\Tr}_{(\Delta,w)} \left[ q^{L_0-c/24} \right] = \frac{q^{\Delta+(2-c)/24}}{\eta(\tau)^2} \,.
\end{equation}

\section{Conclusion}

In this paper we considered the FPL model, with a particular emphasis on the case $\tilde{n} = n^2 -2$ where an extended $W_3$ symmetry is expected in the continuum. 

Through the exact calculation of the partition function on the torus in the continuum limit, the full spectrum was obtained for generic values of $n$ and $\tilde{n}$. These results are consistent with the Coulomb gas approach. In the case $\tilde{n} = n^2 -2$, the spectrum has  been checked numerically by an exact diagonalisation procedure, and an excellent agreement was obtained. 

The integrable structure underlying the FPL model is closely related to a $\mathfrak{su}(3)$ quotient of the Hecke algebra. The $\mathfrak{su}(2)$ case corresponds to the celebrated Temperley-Lieb algebra, which is known to yield the Virasoro algebra in the continuum limit. In the FPL case it is yet to be understood how the $W_3$ algebra emerges from the Hecke algebra. In particular how to realize the zero mode $W_0$ on the lattice, and whether or not it commutes with the transfer matrix, remain open questions which we think are worth investigating. 

Experience in the O$(n)$ model suggests that the FPL model has a rich logarithmic CFT structure in the continuum limit, and understanding the Operator Product Expansions of this theory remains a challenge. Now that the spectrum of this model is under control, it is natural to ask what the fusion rules and structure constants are.

\section*{Acknowledgements} The authors wish to thank Alexi Morin-Duchesne, Raoul Santachiara, Paul Zinn-Justin and Jean-Bernard Zuber for valuable discussions.

\appendix 

\section{Resummation of the loop partition function}
\label{ressumation}

In this appendix we give the derivation of \eqref{loop-partition-function}. The goal is to resum the expression 
\begin{align}
  Z_{\textrm{loop}}\left[ g,e_0,\mathcal{R}\right]  = 
  \text{Vol}(\mathcal{R}) \times
  \sum_{(\vect{q},\vect{q}') \in \mathcal{R}^2}  \exp \left[  2i\pi e_0  (\vect{q}\cdot \vect{\rho}) \wedge (\vect{q}'\cdot \vect{\rho}  ) \right] \, Z_{\vect{q},\vect{q}'}(g)
\end{align}
by Poisson summation and a careful treatment of the $\gcd$ factor, as was done for the O($n$) model \cite{difrancesco_saleur_zuber_coulomb,read_saleur_sigma_01}. 

\subsection{Zero-strand sectors}

Let us start with the sectors where $\vect{q}\cdot \vect{\rho}=0$ (including the purely electric sector where $\vect q=0$). In this case the gcd boils down to $ (\vect{q}\cdot \vect{\rho}) \wedge (\vect{q}'\cdot \vect{\rho}  )=  |\vect{q}'\cdot \vect{\rho}| $ and the sum over $\vect{q}'$ can be Poisson transformed directly. For fixed $\vect q$ one has
\begin{align}
  &\text{Vol}(\mathcal{R}) \times \sum_{\vect{q}' \in \mathcal{R} }
  Z_{\vect{q},\vect{q}'}(g) \exp (2i\pi e_0 \vect{q}'\cdot \vect\rho) \nn \\
  & \qquad\qquad = \sqrt{3} \sum_{\vect{q}' \in \mathcal{R} } Z_{\vect{q},\vect{q}'}(g)  \exp(2i\pi e_0 \vect{q}'\cdot \vect\rho)  \nn \\
  & \qquad\qquad = \frac{g \sqrt{3}}{2\operatorname{Im} \tau |\eta(\tau)|^4}
  \sum_{\vect{q}' \in \mathcal{R} }  \exp\left(-\frac{\pi g}{2 \operatorname{Im} \tau} \left|\vect{q} \tau - \vect{q}' \right|^2 -2i \pi e_0 \vect{q}'\cdot \vect{\rho}\right)  \nn \\
  & \qquad\qquad = \frac{1}{|\eta(\tau)|^4} \sum_{\vect{e} \in \mathcal{R}^*} q^{\delta(\vect{e}- e_0 \vect{\rho},\vect{q})} \bar{q}^{\bar{\delta}(\vect{e}- e_0 \vect{\rho},\vect{q})} \,.
\end{align}

\subsection{Other magnetic sectors} 

In the magnetic sectors  where $\vect{q}\cdot \vect{\rho}\neq 0$, things are a bit more complicated. Following \cite{read_saleur_sigma_01}, we define the function $f_{\vect{q}}: \mathbb{N} \to \mathbb{R}$ as
\begin{equation}
  f_{\vect{q}}(d) = \text{Vol}(\mathcal{R}) \sum_{\vect{q}' \in \mathcal{R} : \, m\wedge m' =d}  Z_{\vect{q},\vect{q}'}(g) \,,
\end{equation}
where the integer $m$ is defined as $m=\vect{q}\cdot \vect{\rho}$ (and $m' =  \vect{q}'\cdot \vect{\rho}$). Note that $f_{\vect{q}}(d) =0$ unless $d |m$ ($d$ is a divisor of $m$). We have 
\begin{align}
  Z_{\textrm{loop}}\left[ g,e_0,\mathcal{R}\right]  =  \sum_{\vect{q} \in \mathcal{R}} \sum_{d >0, d| m}   \cos(2\pi e_0 d)  f_{\vect{q}}(d) \,.
\end{align}
We introduce an auxiliary function $g_{\vect{q}}(d)$ defined as
\begin{align}
  g_{\vect{q}}(d) =  \sum_{d' > 0  : \,  d | d' , \, d' | m} f_{\vect{q}}(d') \,.
\end{align}
Notice that this sum is finite since $m \neq 0$, and that $f_{\vect{q}}(d)$ can be recovered using a variant of the M\"{o}bius inversion formula
\begin{align}
  f_{\vect{q}}(d) =  \sum_{d' > 0  : \,  d | d' , \, d' | m} g_{\vect{q}}(d') \mu (d'/d) \,,
\end{align}
where $\mu$ is the M\"{o}bius function. The quantity $f_{\vect{q}}(d)$ is not easy to calculate directly, because the sum is not over a regular lattice. The auxiliary  $g_{\vect{q}}(d)$ on the other hand is better behaved. Indeed 
\begin{align}
  g_{\vect{q}}(d) =\text{Vol}(\mathcal{R})  \sum_{\vect{q}' \in \mathcal{R} : \, d | m'}  Z_{\vect{q},\vect{q}'} =  \text{Vol}(\mathcal{R}) \sum_{\vect{q}' \in \mathcal{R}_d}  Z_{\vect{q},\vect{q}'} ,\qquad  \mathcal{R}_d = \mathbb{Z} d \vect{e}_1 + \mathbb{Z} (\vect{e}_2 - \vect{e}_1 ) \,.
\end{align}
It is now straightforward to transform this expression, using the Poisson summation formula, yielding
\begin{align}
  g_{\vect{q}}(d)  =  \frac{\text{Vol}(\mathcal{R})}{|\eta(\tau)|^4 \text{Vol}(\mathcal{R}_d)}  \sum_{\vect{e} \in \mathcal{R}_d^*}q^{\delta(\vect{e},\vect{q})} \bar{q}^{\bar{\delta}(\vect{e},\vect{q})}, \qquad  \mathcal{R}_d^* = \mathbb{Z} \vect{\omega}_1  +  \mathbb{Z} \frac{\vect{\omega}_1+\vect{\omega}_2}{d} \,.
\end{align}
where the volume of the unit cell is $\text{Vol}(\mathcal{R}_d) = | d  \vect{e}_1 \wedge (\vect{e}_2- \vect{e}_1) | =  d \sqrt{3}$.  Going back to $f_{\vect{q}}(d)$ we have
\begin{align}
  f_{\vect{q}}(d) =  \frac{1}{|\eta(\tau)|^4}\sum_{d' > 0  : \,  d | d' , \, d' | m}  \frac{\mu(d'/d)}{d'}\sum_{\vect{e} \in \mathcal{R}_{d'}^*}q^{\delta(\vect{e},\vect{q})} \bar{q}^{\bar{\delta}(\vect{e},\vect{q})}
\end{align}
There are duplicates in the sums over $\vect{q}'$, because $\mathcal{R}^*_{d_1} \cap \mathcal{R}^*_{d_2} = \mathcal{R}^*_{d_1 \wedge d_2}$.  We can decompose 
\begin{align}
  \tilde{\mathcal{R}}^*_{d} = \mathcal{R}^*_{d}\setminus  \cup_{d' < d :\, d' | d} \mathcal{R}^*_{d'} =  \left\{  a \vect{\omega}_1 + \frac{b}{d}\vect{\rho} , \, a ,b \in \mathbb{Z}, \,  \gcd(b,d)= 1 \right\}
\end{align}
and
\begin{align}
  \mathcal{R}^*_{d'} =   \sqcup_{k >0:\, k | d'} \tilde{\mathcal{R}}^*_{k}
\end{align}
Therefore
\begin{align}
  f_{\vect{q}}(d) & =  \frac{1}{|\eta(\tau)|^4}\sum_{d' > 0  : \,  d | d' , \, d' | m}  \frac{\mu(d'/d)}{d'}\sum_{k: \, k | d'}\sum_{\vect{e} \in \tilde{\mathcal{R}}_{k}^*}q^{\delta(\vect{e},\vect{q})} \bar{q}^{\bar{\delta}(\vect{e},\vect{q})}\\
  & = \frac{1}{|\eta(\tau)|^4}\sum_{k: \, k | m}   \frac{\mu \left( \frac{k}{d\wedge k} \right)}{m}\frac{\varphi\left( \frac{m}{d} \right)}{\varphi \left( \frac{k}{d \wedge k}\right)}  \sum_{\vect{e} \in \tilde{\mathcal{R}}_{k}^*} q^{\delta(\vect{e},\vect{q})} \bar{q}^{\bar{\delta}(\vect{e},\vect{q})}
\end{align}
where we used the following identity \cite{read_saleur_sigma_01}
\begin{align}
  \sum_{d' > 0  : \,  d | d', \, k | d', \, d' | m}  \frac{\mu(d'/d)}{d'}  =   \frac{1}{m}\sum_{d'' > 0  : \,  \frac{k}{d \wedge k} | d'', \, \, d'' | \frac{m}{d}}  \mu(d'')\frac{m/d}{d''}  =\frac{\mu \left( \frac{k}{d\wedge k} \right)}{m}\frac{\varphi\left( \frac{m}{d} \right)}{\varphi \left( \frac{k}{d \wedge k}\right)} \,,
\end{align}
and $\varphi(n)$ is Euler's totient function. We end up with
\begin{align}
  Z_{\textrm{loop}}\left[ g,e_0,\mathcal{R}\right]  & = \frac{1}{|\eta(\tau)|^4} \left[ \sum_{\substack{ \vect{q} \in \mathcal{R} \\ m = 0 } }  \sum_{\vect{e} \in \mathcal{R}^*} q^{\delta(\vect{e}- e_0 \vect{\rho},\vect{q})} \bar{q}^{\bar{\delta}(\vect{e}-e_0 \vect{\rho},\vect{q})} \right.  \nonumber \\ 
  & + \left. \sum_{\substack{ \vect{q} \in \mathcal{R} \\ m \neq 0 } } \sum_{k | m} \sum_{\vect{e} \in \tilde{\mathcal{R}}_k^*} \Lambda(k,m,e_0) q^{\delta(\vect{e},\vect{q})} \bar{q}^{\bar{\delta}(\vect{e},\vect{q})}   \right] \,,
\end{align}
where
\begin{align}
  \Lambda(k,m,e_0) = \sum_{d | m}  \frac{\mu \left( \frac{k}{d\wedge k} \right)}{m}\frac{\varphi\left( \frac{m}{d} \right)}{\varphi \left( \frac{k}{d \wedge k}\right)} \cos (2\pi d e_0) \,.
\end{align}

\section{Some interesting particular cases}

\subsection{Combinatorial point : $n=\sqrt{2}$ and $\tilde{n}=0$}

The FPL model with $n = \sqrt{2}$ and $\tilde{n} = n^2 -2 =0$ has a vanishing central charge.

This is a combinatorial point \cite{PDiF-PZJ05}, and like in the O$(n)$ model at $n=1$, the coefficients of the eigenvector associated to the ground state have a particular structure, in this case they are all either integers or multiple of $\sqrt{2}$. For example, for $L=6$ (taking into account translation invariance and mirror symmetry) the ground eigenstate is:

\begin{equation}
 \begin{tabular}{| l || r | }
   \hline			
   \looppict{
     \draw (1)to[bend right](6);
     \draw (2)to[bend right](5);
   } & $2$ \\

   \looppict{
     \draw (1)to[bend left](extl2);
     \draw (2)to[bend right](5);
     \draw (6)to[bend right](extr2);
   } &  $12 \sqrt{2}$\\

   \looppict{
     \draw (1)to[bend right](2);
     \draw (5)to[bend right](6);
   } & $18 \sqrt{2}$\\

   \looppict{
     \draw (1)to[bend left](extl1);
     \draw (2)to[bend left](extl2);
     \draw (5)to[bend right](extr2);
     \draw (6)to[bend right](extr1);
   } & $12$\\

   \looppict{
     \draw (1)to[bend left](extl1);
     \draw (2)to[bend right](extr1);
     \draw (5)to[bend right](6);
   } & $8 $\\


   \looppict{
     \draw (1)to[bend right](3);
     \draw (5)to[bend left](extl1);
     \draw (6)to[bend right](extr1);
   } & $5\sqrt{2} $\\

   \looppict{
     \draw (1)to[bend right](3);
     \draw (5)to[bend right](6);
   } & $40 $\\


   \looppict{
     \draw (1)to[bend left](extl2);
     \draw (3)to[bend right](extr2);
     \draw (5) to[bend right](6);
   } & $15\sqrt{2} $\\

   \looppict{
     \draw (5)to[bend left](extl2);
     \draw (6)to[bend right](extr2);
     \draw (2) to[bend right](3);
   } & $20 $\\

   \looppict{
     \draw (2)to[bend right](3);
     \draw (5)to[bend right](6);
   } & $50\sqrt{2} $\\

   \looppict{
     \draw (2)to[bend left](extl1);
     \draw (3)to[bend right](5);
     \draw (6)to[bend right](extr1);
   } & $20\sqrt{2}$\\

   \looppict{
     \draw (2)to[bend left](extl1);
     \draw (3)to[bend left](extl2);
     \draw (5)to[bend right](extr2);
     \draw (6)to[bend right](extr1);
   } & $10$\\
   \hline  
 \end{tabular}\label{eq:2}
\end{equation}

Let us now compute the torus partition function. Setting $\tilde{n}=0$ means $e_0 = 1/4$, and the partition function can be simplified to
\begin{align*}
  Z_{\textrm{loop}}\left[ g,1/4,\mathcal{R}\right]  = \text{Vol}(\mathcal{R}) \sum_{\vect{q},\vect{q}' \in \mathcal{R}}  \cos \left[   \frac{\pi}{2}  (\vect{q}\cdot \vect{\rho}) \wedge (\vect{q}'\cdot \vect{\rho}  ) \right] \, Z_{\vect{q},\vect{q}'}(g)
\end{align*}
This boils down to 
\begin{equation*}
  \begin{split}
    Z_{\textrm{loop}}\left[g,1/4,\mathcal{R}\right] & = \text{Vol}(\mathcal{R})\left(  \sum_{\substack{\vect{q},\vect{q}' \in \mathcal{R} \\ m\wedge m' = 0 \textrm{ mod }4}}  Z_{\vect{q},\vect{q}'} -\sum_{\substack{\vect{q},\vect{q}' \in \mathcal{R} \\ m\wedge m' = 2 \textrm{ mod }4}}  Z_{\vect{q},\vect{q}'}\right)  \\
    &  = \text{Vol}(\mathcal{R}) \left( 2 \sum_{\substack{\vect{q},\vect{q}' \in \mathcal{R} \\ m\wedge m' = 0 \textrm{ mod }4}}  Z_{\vect{q},\vect{q}'} - \sum_{\substack{\vect{q},\vect{q}' \in \mathcal{R} \\ m\wedge m' = 0 \textrm{ mod }2}}  Z_{\vect{q},\vect{q}'}\right) \\
    & = \text{Vol}(\mathcal{R}) \left( 2 \sum_{\vect{q},\vect{q}' \in
        \mathcal{R}_4 } Z_{\vect{q},\vect{q}'}(g) -
      \sum_{\vect{q},\vect{q}' \in \mathcal{R}_2}
      Z_{\vect{q},\vect{q}'}(g)\right)
  \end{split}
\end{equation*}
This means 
\begin{equation*}
    Z_{\textrm{loop}}\left[ g,1/4,\mathcal{R}\right]  = \frac{1}{2} \left(  Z_c[g ,\mathcal{R}_4] - Z_c[g ,\mathcal{R}_2] \right) 
\end{equation*}
This can be further simplified by observing that $\mathcal{R}_{d}$ is a square lattice when $d=2k$ is even. As a consequence the partition function of the two-component compact boson $Z_c[g ,\mathcal{R}_{2k}]$ factorizes into 
\begin{align}
Z_c[g ,\mathcal{R}_{2k}] = Z\left[\sqrt{6g}\right] \, Z\left[ k \sqrt{2g} \right]   
\end{align} 
where $Z[R]$ stands for the partition function of the one-component compact boson at radius $R$ : 
\begin{align}
Z[R] = Z[2/R] =   \frac{1}{|\eta(\tau)|^2}  \sum_{e,q \in \mathbb{Z}} q^{\frac{1}{2}\left( \frac{e}{R} - \frac{q R}{2}\right)^2 } \bar{q}^{\frac{1}{2}\left( \frac{e}{R} + \frac{q R}{2}\right)^2 }
\end{align} 
Setting then $n=\sqrt{2}$, which corresponds to $g = \frac{3}{4}$, we get
\begin{equation*}
  Z_{\textrm{loop}}\left[ 3/4,1/4,\mathcal{R}\right]  = Z\left[ 3\sqrt{2}/2\right] \frac{Z\left[ \sqrt{6}\right] - Z\left[ \sqrt{3/2}\right] }{2} 
 \end{equation*}
As expected the contributions from the identity in both sums cancel out, since the cylinder partition function is trivial. While this theory has a vanishing central charge, the field content is not trivial as illustrated by the partition function on the torus.

\subsection{The model at $n=1$ : symplectic fermions and Dimer model}

\subsubsection{The case $\tilde{n}=1$}

For $n=\tilde{n} = 1$ the FPL model can be mapped to the dimer model, and their partition function must be equal in the continuum. This can be seen from the three-color model, we can choose one of the colour to act as a dimer. Then we obtain the usual partition function of the dimer model on the hexagonal lattice.  We now have $e_0 = \frac{1}{6}$, and therefore
\begin{equation}
  \begin{split}
    Z_{\textrm{loop}}\left[ g,1/6,\mathcal{R}\right]  & = \text{Vol}(\mathcal{R}) \sum_{\vect{q},\vect{q}' \in \mathcal{R}}  \cos \left[   \frac{\pi}{3}  (\vect{q}\cdot \vect{\rho}) \wedge (\vect{q}'\cdot \vect{\rho}  ) \right] \, Z_{\vect{q},\vect{q}'}(g) \\
    & = \text{Vol}(\mathcal{R}) \left( \sum_{\substack{\vect{q},\vect{q}' \in \mathcal{R} \\ m\wedge m' = 0 \textrm{ mod }6}}  Z_{\vect{q},\vect{q}'}  + \frac{1}{2} \sum_{\substack{\vect{q},\vect{q}' \in \mathcal{R} \\ m\wedge m' = \pm 1 \textrm{ mod }6}}  Z_{\vect{q},\vect{q}'} \right.  \\
    &  \left. - \frac{1}{2} \sum_{\substack{\vect{q},\vect{q}' \in \mathcal{R} \\ m\wedge m' = \pm 2 \textrm{ mod }6}}  Z_{\vect{q},\vect{q}'} -  \sum_{\substack{\vect{q},\vect{q}' \in \mathcal{R} \\ m\wedge m' = 3 \textrm{ mod }6}}  Z_{\vect{q},\vect{q}'} \right) \\
    &  = \text{Vol}(\mathcal{R}) \left( 3 \sum_{\substack{\vect{q},\vect{q}' \in \mathcal{R} \\ m\wedge m' = 0 \textrm{ mod }6}}  Z_{\vect{q},\vect{q}'} - \frac{3}{2} \sum_{\substack{\vect{q},\vect{q}' \in \mathcal{R} \\ m\wedge m' = 0 \textrm{ mod }3}}  Z_{\vect{q},\vect{q}'}  \right.     \\ & \left.-  \sum_{\substack{\vect{q},\vect{q}' \in \mathcal{R} \\ m\wedge m' = 0 \textrm{ mod }2}}  Z_{\vect{q},\vect{q}'} + \frac{1}{2} \sum_{\vect{q},\vect{q}' \in \mathcal{R}}  Z_{\vect{q},\vect{q}'}\right) \\
    & = \frac{1}{2} \left(  Z_c[g, \mathcal{R}_6] - Z_c[g, \mathcal{R}_3]  - Z_c[g, \mathcal{R}_2] +  Z_c[g, \mathcal{R}] \right) 
  \end{split}
\end{equation}
For $g=2/3$ we again get the cancellation from $Z_c[2/3, \mathcal{R}_3] = Z_c[2/3, \mathcal{R}]$, and we have 
\begin{equation}
  Z_{\textrm{loop}}\left[2/3,1/6,\mathcal{R}\right] =  \frac{1}{2} \left(  Z_c[2/3, \mathcal{R}_6]  - Z_c[2/3, \mathcal{R}_2] \right) 
  \label{eq:partition_function_c-2}
\end{equation}
Since $ \mathcal{R}_6$ and $ \mathcal{R}_2$ are square lattices, we get a factorization
\begin{align}
  Z_{\textrm{loop}}\left[2/3,1/6,\mathcal{R}\right] =    Z[1]    \, \frac{ \left(  Z[\sqrt{12}]  - Z[\sqrt{3}] \right) }{2}
\end{align}  
Finally the term $(Z[\sqrt{12}]  - Z[\sqrt{3}])/2 =1$ since it is the partition function of the $c=0$ minimal model $M(2,3)$, and we have 
\begin{align}
  Z_{\textrm{loop}}\left[2/3,1/6,\mathcal{R}\right] =    Z[1]   
\end{align}   
which is the partition function of the one-component compact boson at radius $R=1$, or equivalently the one of the Dirac fermion. This is also the partition function of the Dimer model.  
\begin{equation}
   Z_{\textrm{loop}}\left[2/3,1/6,\mathcal{R}\right] =   Z_{\textrm{Dimer}} = \frac{1}{2} \left( \left|\frac{\theta_2(0|
          \tau)}{\eta(\tau)}\right|^2 + \left|\frac{\theta_3(0|
          \tau)}{\eta(\tau)}\right|^2 + \left|\frac{\theta_4(0|
          \tau)}{\eta(\tau)}\right|^2 \right)
\end{equation}

\subsubsection{The case $\tilde{n}= -1$}

When $n=1$ the $W_3$ symmetry is expected for $\tilde{n} = n^2 -2 = -1$. This means $c=-2$ ($g = \frac{2}{3}$ and $e_0 = \frac{1}{3}$).  The central charge $c=-2$ does not correspond to the dimer model ($c=1$), because of the behaviour of the non-contractible loop. It's the central charge of the symplectic fermions model \cite{gaberdiel96} (interestingly the $\mathcal{W}_3$ algebra is is a sub-algebra of the symplectic fermions ones). It's a different take on the well known correspondence between dimers and symplectic fermions \cite{duchesne_dimer_symp}. The partition function vanishes identically in this case. 

\begin{equation*}
  \begin{split}
    Z_{\textrm{loop}}\left[ g,1/3,\mathcal{R}\right]  & = \text{Vol}(\mathcal{R}) \sum_{\vect{q},\vect{q}' \in \mathcal{R}}  \cos \left[   \frac{2\pi}{3}  (\vect{q}\cdot \vect{\rho}) \wedge (\vect{q}'\cdot \vect{\rho}  ) \right] \, Z_{\vect{q},\vect{q}'}(g) \\
    & = \text{Vol}(\mathcal{R}) \left( \sum_{\substack{\vect{q},\vect{q}' \in \mathcal{R} \\ m\wedge m' = 0 \textrm{ mod }3}}  Z_{\vect{q},\vect{q}'} - \frac{1}{2}\sum_{\substack{\vect{q},\vect{q}' \in \mathcal{R} \\ m\wedge m' = \pm 1 \textrm{ mod }3}}  Z_{\vect{q},\vect{q}'} \right) \\
    &  = \text{Vol}(\mathcal{R}) \left( \frac{3}{2} \sum_{\substack{\vect{q},\vect{q}' \in \mathcal{R} \\ m\wedge m' = 0 \textrm{ mod }3}}  Z_{\vect{q},\vect{q}'} - \frac{1}{2}\sum_{\vect{q},\vect{q}' \in \mathcal{R}}  Z_{\vect{q},\vect{q}'}\right) \\
    & = \frac{1}{2} \left( Z_c[g, \mathcal{R}_3] - Z_c[g, \mathcal{R}] \right)
  \end{split}
\end{equation*}
However for $g=2/3$, one finds that $Z_c[2/3, \mathcal{R}_3] = Z_c[2/3, \mathcal{R}]$, so we get
\begin{equation*}
  Z_{\textrm{loop}}\left[ g=2/3,e_0=1/3,\mathcal{R}\right] =0\label{eq:4}
\end{equation*}

\bibliographystyle{unsrt}
\bibliography{biblio}

\end{document}